\documentclass[iop,numberedappendix,twocolappendix]{emulateapj}

\newcommand{\cha}{\textit{Chandra\/}}
\def\xmm{{XMM-{\it Newton\/}}}
\def\XMM{{XMM-{\it Newton\/}}}

\def\swi{{{\it Swift}-BAT}\/}
\def\xrt{{{\it Swift}-XRT}\/}

\def\nustar{{\it NuSTAR}}
\def\nus{{\it NuSTAR}}
\def\NuSTAR{{\it NuSTAR}}

\def\myt{\texttt{MYTorus}}
\def\borus{\texttt{borus02}}

\shorttitle{\nustar\ and \xmm\ spectral analysis of three ``soft--$\Gamma$'' candidate CT-AGNs}
\shortauthors{Marchesi et al.}

\usepackage{natbib}
\usepackage{float}
\usepackage{color}
\usepackage{graphicx}
\usepackage{gensymb}
\usepackage{array}
\usepackage{mathtools}
\usepackage{multirow}
\usepackage{enumitem}
\usepackage{longtable}
\usepackage{hyperref}

\begin{document}

\title{Compton-thick AGN in the \nustar\ era. V: Joint \nustar\ and \xmm\ spectral analysis of three ``soft--$\Gamma$'' candidate CT-AGN\MakeLowercase{s} in the \swi\ 100-month catalog}

\author{S. Marchesi\altaffilmark{1}, M. Ajello\altaffilmark{1}, X. Zhao\altaffilmark{1}, A. Comastri\altaffilmark{2}, V. La Parola\altaffilmark{3}, A. Segreto\altaffilmark{3}}

\altaffiltext{1}{Department of Physics and Astronomy, Clemson University,  Kinard Lab of Physics, Clemson, SC 29634, USA}
\altaffiltext{2}{INAF - Osservatorio di Astrofisica e Scienza dello Spazio di Bologna, Via Piero Gobetti, 93/3, 40129, Bologna, Italy}
\altaffiltext{3}{INAF - Istituto di Astrofisica Spaziale e Fisica Cosmica, Via U. La Malfa 153, I-90146 Palermo, Italy}

\begin{abstract}
We present the joint \nus\ and \xmm\ spectral analysis in the 0.6--70\,keV band of three candidate Compton thick (CT--) AGN selected in the 100-month \swi\ catalog. These objects were previously classified as CT--AGNs based on low quality \xrt\ and \swi\ data, and had soft photon indices ($\Gamma$$>$2.2) that suggested a potential overestimation of the line of sight column density ($N_{\rm H, l.o.s.}$). Thanks to the high-quality \nus\ and \xmm\ data we were able to determine that in all three objects the photon index was significantly overestimated, and two out of three sources are reclassified from CT to Compton thin, confirming a previously observed trend, i.e., that a significant fraction of BAT-selected, candidate CT-AGNs with poor soft X-ray data are reclassified as Compton thin when the \nus\ data are added to the fit. 
Finally, thanks to both the good \xmm\ spatial resolution and the high \nus\ and \xmm\ spectral quality, we found that the third object in our sample was associated to the wrong counterpart: the correct one, 2MASX J10331570+5252182, has redshift $z$=0.14036, which makes it one of the very few candidate CT-AGNs in the 100-month BAT catalog detected at $z$$>$0.1, and a rare CT quasar.
\end{abstract}


\section{Introduction}
The study in the X-rays of heavily obscured active galactic nuclei (AGN) and, more specifically, of the obscuring material commonly defined as ``torus'' and constituted of dust and cold molecular gas surrounding the accreting supermassive black holes (SMBHs) in the center of the galaxies, has been significantly improved by the launch of the Nuclear Spectroscopic Telescope Array \citep[hereafter \nustar, ][]{harrison13}. In fact, \nustar\ has excellent sensitivity over the 3--78\,keV range and is  the first telescope with focusing optics at $>$10\,keV: this makes it an ideal instrument to study heavily obscured AGNs, since their observed X-ray emission peaks at $\sim$30--50\,keV \citep[see, e.g.,][]{antonucci93,comastri95,gilli07,ajello08a}, in the so-called ``Compton hump'', while little to no photons at energies below 5\,keV escape the obscuring material \citep[see, e.g.,][]{murphy09,brightman11,koss16}.

During the years, \nustar\ targeted many well-known CT-AGNs in the nearby Universe ($z$$\leq$0.1), making it possible to characterize them with unprecedented accuracy \citep[see, e.g.,][]{balokovic14,puccetti14,annuar15,bauer15,brightman15,koss15,rivers15,masini16,puccetti16,ursini18,zhao19b,zhao19a}, using Monte Carlo radiative transfer codes specifically developed to fit the X-ray spectra of heavily obscured AGN  \citep[e.g.,][]{ikeda09,murphy09,yaqoob10,brightman11,yaqoob12,liu14,furui16,balokovic18}. These models are more refined than simple phenomenological ones, and allow one to measure important parameters, such as the torus covering factor, $f_c$, and the torus average column density, $N_{\rm H, tor}$. To be used effectively, however, these models require an excellent spectral statistics in the 2--50\,keV band, a condition that cannot be satisfied neither by one of the several 0.3--10\,keV facilities nor by \swi.

In \citet[][M18, M19 hereafter]{marchesi18,marchesi19} we analyzed the combined 2--100\,keV spectra of 35 candidate nearby (average redshift $\langle$$z$$\rangle$=0.03) CT-AGN, i.e., all the 100-month BAT candidate CT-AGN having archival \nus\ data. 2--10\,keV data have been obtained using archival \xmm, \cha\ and \xrt\ data. We discovered a systematic trend to artificially overestimate the line of sight (l.o.s). column density and the steepness of the spectrum when only the 2--10\,keV and the \swi\ data are included in the fit. This effect is variability-- and model--independent and stronger in sources with low statistics (net cts$<$100) in the 0.3--10\,keV band, i.e., mostly objects with only a \xrt\ or a short ($<$10\,ks) \cha\ observation available. In these objects, the l.o.s. column density is overestimated, on average, by $\sim$45\%, while the average photon index variation is $\Delta$$\Gamma$=0.25. No significant trend is instead observed in sources with deep ($\geq$20\,ks) \xmm\ observations. As a consequence, only about half (54$^{+10}_{-13}$\%) of the candidate nearby CT-AGN already reported in the literature are confirmed as bona-fide CT-AGN. 

This result has important implications for our understanding of the CXB and the total accretion history of the Universe. For example, based on our results the observed CT-AGN fraction in the 70-month \swi\ catalog \citep[7.6$^{+1.1}_{-2.1}$\%][]{ricci15}, decreases to 6.0$^{+0.4}_{-0.5}$\% and potentially even down to $\sim$4\%, extrapolating the results of our work to the population of candidate CT-AGN with no \nus\ data available. Notably, the low-$z$ observed and intrinsic CT-AGN fractions play an important role in supermassive black hole population synthesis and CXB models \citep[see, e.g.,][]{gilli07,treister09,ballantyne11,ueda14,ananna19}, leaving the total contribution of CT-AGN to the CXB still debated.

Within the framework of a broader project in which we aim to characterize the whole CT-AGN population of nearby, \swi--selected candidate CT-AGNs, in this work we therefore study the joint \nus--\xmm\ spectrum of the three candidate CT-AGNs selected in the 100-month BAT catalog (Marchesi et al. in prep.) having spectra with less than 35 counts in the 2--10\,keV band and best-fit $\Gamma\geq$2.2. 

The paper is organized as follows: in Section \ref{sec:sample} we present the sample used in this work and we describe the data reduction and spectral extraction process for both \nustar\ and the 0.3--10\,keV observations. In Section \ref{sec:model} we describe the models used to perform the spectral fitting. In Section \ref{sec:results} we report the results of the spectral analysis and an extended discussion on the re-association of one of the three sources in our sample.
Finally, we discuss our results and report our conclusions in Section \ref{sec:concl}. All reported errors are at a 90\% confidence level, if not otherwise stated.

\section{Sample selection and data reduction}\label{sec:sample}
The three sources analyzed in this work (ESO 244-IG 030, $z$=0.0256; ESO 317-G 041, $z$=0.0193; and 2MASX J10331570+5252182, $z$=0.14036) have been selected from the Palermo BAT 100-month catalog\footnote{\url{http://bat.ifc.inaf.it/100m\_bat\_catalog/100m\_bat\_catalog\_v0.0.htm}}
The data have been processed with the BAT\_IMAGER code \citep{segreto10}, and the spectra are background subtracted and exposure-averaged. We use the standard \swi\ spectral redistribution matrix\footnote{Available at \url{http://heasarc.gsfc.nasa.gov/docs/heasarc/caldb/data/swift/\\bat/index.html}}.  

All the sources are reported to be candidate CT-AGNs \citep{ricci15}, but they also have a particularly soft best-fit photon index ($\Gamma$$\geq$2.2, see Table \ref{tab:sample}): while similar photon indices are not necessarily unphysical, the typical AGN photon index is usually significantly harder \citep[$\Gamma$$\sim$1.4--2.0; see, e.g.,][]{nandra94,risaliti04,ueda14,marchesi16c}. Furthermore, given the known degeneracy between $\Gamma$ and the line of sight column density, $N_{\rm H, z}$, if the photon index value turns out to be overestimated, it is likely that $N_{\rm H, z}$ could also be overestimated, potentially moving these sources from the Compton thick to the Compton thin regime. Notably, all three sources had only been observed in the 0.5--10\,keV band with \xrt\ and have poor count statistics (15--25 net counts; see Table \ref{tab:sample}), thus $\Gamma$ and $N_{\rm H, z}$ both had relatively large (30-40\%) uncertainties.  

To properly assess the X-ray spectral properties of these three AGN we proposed for a joint, 20\,ks \xmm, 30\,ks \nustar\ follow-up: our proposal was accepted (\nustar\ GO Cycle 4, proposal ID: 4253; PI S. Marchesi) and the observations took place between May and June 2018. We report a summary of the observations in Table \ref{tab:sample}.

The \xmm\ data was reduced using the SAS v16.1.0\footnote{\url{http://xmm.esa.int/sas}} \citep{jansen01} packages and adopting standard procedures. We extracted the source spectra from a 15$^{\prime\prime}$ circular region: if the source is observed on-axis, as it is the case for the three sources studied in this work, such a radius is equivalent to $\sim$70\% of the encircled energy fraction at 5\,keV for all the three \xmm\ 0.5--10\,keV cameras (MOS1, MOS2 and pn). The background spectra were instead extracted from a circular region having radius $r$=45$^{\prime\prime}$: for all sources, the background spectra were extracted from a part of the CCD  located near the source and not contaminated by other objects. Finally, each spectrum has been binned with at least 20 counts per bin.

\begin{figure}
  \centering
  \includegraphics[width=0.45\textwidth]{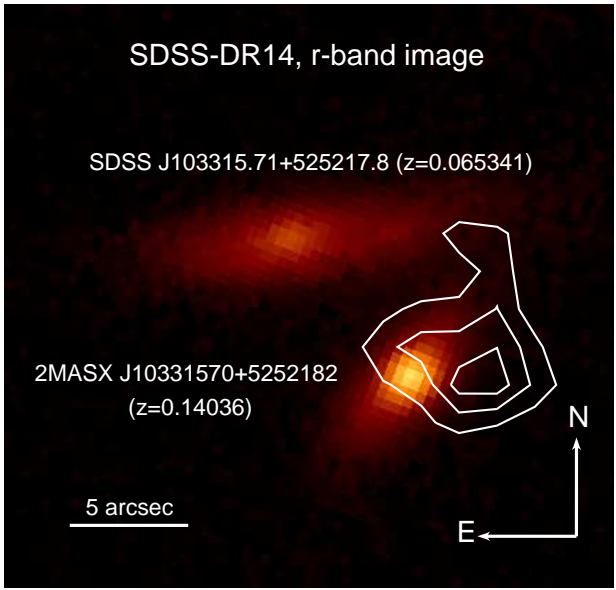}
\caption{\small SDSS-DR14 r-band image at the position of 4BPCJ1033.4+5252, with the 0.5--10\,keV \xmm\ pn confidence contours overlapped. As can be seen, the X-ray emission is originated by the Southern source, 2MASX J10331570+5252182. As a reference, the 4PBC source 95\,\% confidence position uncertainty is $r$=3.1$^{\prime}$.}\label{fig:J1033_rband}
\end{figure}

We point out that the \swi\ source 4BPCJ1033.4+5252 \citep[i.e., source SWIFTJ1033.8+5257 in ][]{ricci15} was originally associated to the galaxy SDSS J103315.71+525217.8 at $z$=0.0653, a source located at $\sim$10$^{\prime\prime}$ from 2MASX J10331570+5252182. As can be seen in Figure \ref{fig:J1033_rband}, however, the \xmm\ centroid is closer to the Southern source, i.e., 2MASX J10331570+5252182 ($z$=0.14036), thus supporting the hypothesis that this is instead the correct counterpart. We will further strengthen this claim in Section \ref{sec:J1033}, where we report the joint \nustar\ and \xmm\ analysis of 4BPCJ1033.4+5252.

Finally, for all the objects the data retrieved for both  \nustar\  Focal Plane Modules \citep[FPMA and FPMB;][]{harrison13} were processed using the  \nustar\  Data Analysis Software (NUSTARDAS) v1.5.1. 
The event data files were calibrated running the {\tt nupipeline} task using the response file from the Calibration Database (CALDB) v. 20181030. With the {\tt nuproducts} script we generated both the source and background spectra, and the ancillary and response matrix files. 
For both focal planes, we used a circular source extraction region with a radius chosen to maximize the spectral signal-to-noise ratio\footnote{40$^{\prime\prime}$ for ESO 244-IG 030, 60$^{\prime\prime}$ for ESO 317-G 041, 50$^{\prime\prime}$ for 2MASX J10331570+5252182 and 60$^{\prime\prime}$ for 2MASX J10313591-4206093, which we analyze in the Appendix.} (S/N), and centered on the target source; for the background we used the same extraction region positioned far from any source contamination in the same frame. The \nustar\ spectra have then been grouped with at least 20 counts per bin, and cover the energy range from 3 to 50--70\,keV, depending on the quality of the data. 

\begingroup
\renewcommand*{\arraystretch}{1.15}
\begin{table*}
\centering
\scalebox{0.83}{
\begin{tabular}{lccccccccccccc}
\hline
\hline
  \multicolumn{1}{c}{4PBC ID} & \multicolumn{1}{c}{Source name} &   \multicolumn{1}{c}{R.A.} & \multicolumn{1}{c}{Decl} & \multicolumn{1}{c}{$z$} & \multicolumn{1}{c}{$N_{\rm H, z,Swi}$} &\multicolumn{1}{c}{$\Gamma_{Swi}$} & \multicolumn{1}{c}{cts} & \multicolumn{1}{c}{Telescope} & \multicolumn{1}{c}{ObsID} & \multicolumn{1}{c}{Date} &  \multicolumn{1}{c}{Exp} &  \multicolumn{1}{c}{Rate}\\
  & & deg & deg & & & & & & & & ks & cts s$^{-1}$\\
  \multicolumn{1}{c}{(1)} & (2) & (3) & (4) & (5) & (6) & (7) & (8) & (9) & (10) & (11) & (12) & (13)\\
\hline
\multirow{2}{*}{J0129.8-4218} & \multirow{2}{*}{ESO 244-IG 030} &  \multirow{2}{*}{22.46346} & \multirow{2}{*}{-42.32647} &  \multirow{2}{*}{0.0256} & \multirow{2}{*}{24.2$_{-0.2}^{+0.4}$} & \multirow{2}{*}{2.45$^{+0.35}_{-0.40}$} & \multirow{2}{*}{15} & XMM & 0830500101 & 2018-05-23 & 66.6 & 0.018\\ 
& & & & & & & & \nustar & 60468001002 & 2018-05-23 & 60.1 & 0.015\\
\multirow{2}{*}{J1031.5-4203} & \multirow{2}{*}{ESO 317-G 041} &       \multirow{2}{*}{157.84633}  & \multirow{2}{*}{-42.06061}  &  \multirow{2}{*}{0.0193} & \multirow{2}{*}{24.3$_{-0.2}^{+0.4}$} & \multirow{2}{*}{2.20$^{+0.27}_{-0.22}$} & \multirow{2}{*}{12} & XMM & 0830500201 & 2018-05-27 & 75.8 & 0.003\\%
& & & & & & & &\nustar & 60468002002 & 2018-05-27 & 61.3 & 0.030\\
\multirow{2}{*}{J1033.4+5252} & \multirow{2}{*}{2MASX J10331570+5252182} & \multirow{2}{*}{158.31548}  & \multirow{2}{*}{52.87164} & \multirow{2}{*}{0.14036} & \multirow{2}{*}{24.3$_{-0.2}^{+0.3}$} & \multirow{2}{*}{2.35$^{+0.24}_{-0.30}$} & \multirow{2}{*}{24} & XMM & 0830500301 & 2018-06-02 & 80.7 & 0.006 \\
& & & & & & & &\nustar & 60468003002 & 2018-06-02 & 61.0 & 0.012\\
\hline
\hline
\end{tabular}}\caption{\small \raggedright Properties of the candidate CT-AGNs analyzed in this work. Column (1): ID from the Palermo BAT 100-month catalog (Cusumano et al. 2019 in prep.). (2): source name. (3) and (4): right ascension and declination (J2000 epoch). (5): redshift. (6): line of sight column density from the joint \xrt--\swi\ spectral fitting. (7): photon index from the joint \xrt--\swi\ spectral fitting. (8): net counts in the 0.5--10\,keV archival \xrt\ spectrum.
 (9): telescope used in the analysis. (10): observation ID. (11): observation date. (12): total exposure, in ks. For \xmm\ and \nustar, this is the sum of the exposures of each camera. (13): average count rate (in cts s$^{-1}$), weighted by the exposure for \xmm\ and \nustar, where observations from multiple instruments are combined. Count rates are computed in the 3--70\,keV band for \nustar\ and in the 2--10\,keV band otherwise.}\label{tab:sample}
\end{table*}
\endgroup

\section{Spectral fitting procedure}\label{sec:model}
The spectral fitting procedure was performed using the XSPEC software \citep{arnaud96}; the Galactic absorption values is the one measured by \citet{kalberla05}. We used \citet{anders89} cosmic abundances, fixed to the solar value, and the \citet{verner96} photoelectric absorption cross-section. The data is fitted between 0.6 and 70\,keV in ESO 244-IG 030 and 2MASX J10331570+5252182, while in ESO 317-G 041 we select the 3--70\,keV energy range to avoid an artificial flattening in the best-fit photon index (see Section \ref{sec:eso317} for further details).

The spectra are fitted using two Monte Carlo radiative transfer codes specifically developed to characterize the spectra of heavily obscured AGNs: \myt\ \citep{murphy09,yaqoob12,yaqoob15}, used both in its ``coupled'' and in its ``decoupled'' configuration, and \borus\ \citep{balokovic18}. 

\subsection{\myt}\label{sec:myt}
In \myt, the obscuring material is toroidally shaped and azimuthally symmetric: the torus half-opening angle is fixed to $\theta_{\rm OA}$=60$\degree$ (i.e., the torus covering factor is fixed to $f_c$=cos($\theta_{\rm OA}$)=0.5), while the angle between the torus axis and the observer is a free parameter, which varies in the range $\theta_{\rm obs}$=[0--90]$\degree$.

The \myt\ model is divided in three distinct components, to study the properties of an obscured AGN in a self-consistent way. The first one is a multiplicative component and is applied to the main power law continuum: it describes the photoelectric absorption and Compton scattering attenuation, and allows one to measure the neutral hydrogen equatorial column density ($N_{\rm H,eq}$). The second component is defined as the ``reprocessed component'', or scattered continuum, and models the photons that reach the observer after one or more interactions with the obscuring material nearby the accreting supermassive black hole.  We refer to the normalization of the reprocessed component with respect to the main continuum as A$_{\rm S}$.  The cutoff energy of this second component can vary in the range of $E\rm_{c}$ = [160--500]\,keV: in this analysis, we fix the parameter to a typical value, $E\rm_{c}$ = 500\,keV, since in none of the three sources analyzed in this work we find a significant improvement in the fit when leaving $E\rm_{c}$ free to vary. 
Finally, the last component models the emission of the Iron fluorescent lines commonly observed in heavily obscured AGN, namely, the Fe K$\alpha$ line at 6.4\,keV and the K$\beta$ line at 7.06\,keV. The normalization of these lines is with respect to the main continuum is here named A$_{\rm L}$ and is tied to A$_{\rm S}$, since both components are expected to have the same origin.

In \texttt{XSPEC} the fit to a spectrum with \myt\ is described as follows:
\begin{equation}
\label{eq:coupled}
\begin{aligned}
Model_{MyT}=&constant_1*phabs*\\
&(mytorus\_ Ezero\_v00.fits*pow_1\\
&+A_S*mytorus\_scatteredH500\_v00.fits\\
&+A_L*mytl\_V000010nEp000H500\_v00.fits\\
&+constant_2*pow_2+mekal)\\
\end{aligned}
\end{equation}
where $constant_1$ is the cross-normalization constant $C_{Ins}$ between the \xmm\ and the \nus\ data, $phabs$ is the Galactic absorption, the table \textit{mytorus\_Ezero\_v00.fits} models the absorption to the zeroth-order continuum, $pow_1$, \textit{mytorus\_scatteredH500\_v00.fits} accounts for the reprocessed continuum, \textit{mytl\_V000010nEp000H500\_v00.fits} models the Iron K$\alpha$ and K$\beta$ lines, and $constant_2$ accounts for the fraction of emission scattered, rather than absorbed by the obscuring torus (the photon index of this second power law component, $pow_2$, is fixed to the one of $pow_1$). Finally, in 2MASX J10331570+5252182 we add to the model a phenomenological \texttt{mekal} component \citep{mewe85} to account for the soft excess observed below 1\,keV.

In its standard, ``coupled'' configuration, the column density of the reprocessed emission component is tied to the one associated to the zeroth-order continuum, and $\theta_{\rm obs}$ is left free to vary. ``\myt\ decoupled'' \citep{yaqoob12}, instead, allows one to separate the l.o.s. column density, $N_{\rm H,l.o.s.}$, from the reprocessed component column density, $N_{\rm H,S}$, which can also be treated as a measurement of the torus average column density, since the reprocessed emission is given by photons scattered into the observer l.o.s. from all possible directions. In ``\myt\ decoupled'', the viewing angle of the zeroth-order continuum absorber is fixed to $\theta_{\rm obs}$=90$\degree$, so that $N_{\rm H,eq}$=$N_{\rm H,l.o.s.}$; the viewing angle for the reprocessed component can instead be either  $\theta_{\rm obs, AS, AL}$=90$\degree$ (depicting a scenario where most of the reprocessed emission comes from material located between the observer and the accreting SMBH) or $\theta_{\rm obs, AS, AL}$=0$\degree$ (where most of the reprocessed emission comes from the back side of the torus). In all the three sources studied in this work, we find that the  $\theta_{\rm obs, AS, AL}$=90$\degree$ scenario is preferred by the data.

\subsection{\borus}\label{sec:borus}
The \borus\ \citep{balokovic18} radiative transfer code models the reprocessed emission component of an AGN X-ray spectrum, i.e., following the \myt\ nomenclature we introduced in the previous section, the ``reprocessed component'' and the neutral Fe emission lines. The obscuring material geometry in \borus\ is quasi--toroidal, with conical polar cutouts. The torus covering factor, $f_c$, is a free parameter of the model and can vary in the range $f_c$=[0.1--1]. The angle between the torus axis and the observer can also vary, but we fix it to $\theta_{\rm obs}$=87\degree, i.e., the upper boundary of the parameter in the model, corresponding to an almost ``edge-on'' configuration, to reduce potential degeneracies between this parameter and $f_c$\footnote{We point out, however, that we are working on a paper on $\theta_{\rm obs}$ and its possible trends with other spectral parameters (X. Zhao et al. 2019, in prep.), and we find that leaving $\theta_{\rm obs}$ free to vary does not significantly affect the measurement of the other parameters.}. 

Another free parameter in \borus\ is the average torus column density ($N_{\rm H, tor}$): in this work, to find the best-fit $N_{\rm H, tor}$ we follow the approach adopted by \citet{balokovic18} when fitting single-epoch \nustar\ observations. We thus fit each of our spectra 36 times, each time fixing $N_{\rm H, tor}$ to a different value in the range Log($N_{\rm H, tor}$)=[22--25.5] (i.e., the lower and upper boundaries of the parameter in \borus); in each iteration, we increase  the Log($N_{\rm H, tor}$) value by 0.1. The best-fit $N_{\rm H, tor}$ is then the one corresponding to the fit having the minimum $\chi^2$. The best fit $f_c$ is then the one obtained at this $N_{\rm H, tor}$. 

The \texttt{XSPEC} configuration of \borus\ is:
\begin{equation}
\label{eq:Borus}
\begin{aligned}
Model_{bor}=&constant_1*phabs*(borus02\_v170323a.fits\\
&+zphabs*cabs*cutoffpwl_1+constant_2*\\
&cutoffpwl_2+mekal) 
\end{aligned}
\end{equation}
where $borus02\_v170323a.fits$ is an additive table that models the reprocessed components, i.e., both the reprocessed continuum and the fluorescent lines. Since \borus\ does not model the l.o.s. absorption, we describe it with the components \texttt{zphabs $\times$ cabs}, which properly treats Compton scattering losses out of the line of sight: the free parameter associated to these two components is $N_{\rm H, l.o.s.}$, which is identical in \texttt{zphabs} and \texttt{cabs} and varies independently from $N_{\rm H, tor}$; finally, $cutoffpwl_1$ and $cutoffpwl_2$ are two power law components with high-energy cutoff at E=500\,keV, for consistency with \myt. The other components in the model are the same used with \myt.

\section{Results}\label{sec:results}
In Table \ref{tab:results} we report a summary of the best-fit parameters obtained using the models described in the previous section. In the next sections, we will describe in detail the results of the spectral analysis for each of the three objects in our sample.

\subsection{ESO 244-IG 030}\label{sec:eso244}
We first fit the ESO 244-IG 030 joint \nus--\xmm\ spectrum using \myt\ in its ``coupled'' configuration, obtaining a good best fit statistic ($\chi^2$/d.o.f.=104.4/97).
We measure a photon index best-fit $\Gamma$=1.86$_{-0.17}^{+0.16}$, a value close to the typical AGN one and harder than the one reported in \citet{ricci15} using the \xrt\ and \swi\ data alone ($\Gamma$=2.45$^{+0.35}_{-0.40}$). We also measure a significantly different line-of-sight (l.o.s.) column density measurement: based on the joint \nus--\xmm\ fit, ESO 244-IG 030 is only mildly obscured, having Log(N$\rm _{H,l.o.s}$)=22.75$_{-0.08}^{+0.08}$.

To confirm the result obtained using \myt\ coupled, we refit our data using \myt\ in its ``decoupled'' configuration, allowing the column density of the reprocessed component (N$\rm _{H,S}$) to vary independently from the the l.o.s. column density. We find that the two models are statistically identical ($\chi^2$/d.o.f.=104.4/97 for \myt\ coupled and $\chi^2$/d.o.f.=104.2/97 for \myt\ decoupled) and all the free parameters measurements are in excellent agreement. Since the contribution of the reprocessed component is negligible, as expected for sources with low obscuration, N$\rm _{H,S}$ is unconstrained. We report the best-fit \myt\ decoupled model in Figure \ref{fig:spectra}, top left panel. 

Finally, we fit the ESO 244-IG 030 spectrum using \borus (Figure \ref{fig:spectra}, top right panel): once again, the best-fit model is statistically identical to the other two ($\chi^2$/d.o.f.=104.2/97) and all the main fit parameters are in agreement with those computed using \myt\ in either configuration. Following the approach described in \ref{sec:borus}, we find that the best-fit average torus column density is Log(N$\rm _{H,tor}$)=22.2: we show the spectrum and best-fit model in Figure \ref{fig:spectra}, top right panel. The best-fit covering factor corresponding to this N$\rm _{H,tor}$ is $f_c$=1.00, but the parameter is basically unconstrained: in fact, fixing the covering factor to $f_c$=0.1 (i.e., the parameter lower limit) we obtain a best-fit statistic $\chi^2$/d.o.f.=104.4/97, with a statistic variation $\Delta$$\chi^2$=0.2. Once again, this is not an unexpected result, since according to our model the contribution of the reprocessed component to the 2--70\,keV emission is negligible.

In conclusion, we find that ESO 244-IG 030 is not a Compton thick AGN, and is in fact only moderately obscured, fully supporting our original assumption (i.e., that in sources with only low-quality 2--10\,keV and \swi\ data and soft best-fit photon index the l.o.s. column density is likely overestimated). 

\subsection{ESO 317-IG 030}\label{sec:eso317}
Due to the low quality of the \xmm\ data below 3\,keV, which resulted in an artificial under-estimation of the photon index $\Gamma$, we decided to fit the spectrum of ESO 317-IG 030 in the 3--70\,keV energy range.

Following the same approach described in the previous section, we first fit the joint  \nus--\xmm\ spectrum using \myt\ in its ``coupled'' configuration: this first model leads to an excellent fit statistic ($\chi^2$/d.o.f.=118.4/122). The photon index best-fit value we obtain is significantly harder than the one measured using the \xrt\ and \swi\ data ($\Gamma_{\rm XRT-BAT}$=2.20$^{+0.27}_{-0.22}$), i.e., $\Gamma_{\rm XMM-NuS}$=1.47$_{-0.07}^{+0.25}$. Similarly to what we observed in ESO 244-IG 030, this change implies a significant variation in the l.o.s column density measured with \nus\ and \xmm\ with respect to the one derived from the \xrt\ and \swi\ data: indeed, we find that N$\rm _{H,l.o.s,XMM-NuS}$ is three times smaller than N$\rm _{H,l.o.s,XRT-BAT}$ and ESO 317-IG 030 is consequently reclassified as a Compton thin AGN, having Log(N$\rm _{H,l.o.s}$)= 23.88$_{-0.05}^{+0.08}$.

We then re-fit the data with \myt\ in its decoupled configuration: we show the best fit model in Figure \ref{fig:spectra}, central left panel. The best fit statistic is equivalent to the one of  \myt\ coupled ($\chi^2$/d.o.f.=117.9/121): similarly, the best fit parameters are also in close agreement with those obtained using ``\myt\ coupled'', the best-fit photon index being $\Gamma$=1.60$_{-0.20}^{+0.34}$ and the l.o.s. column density being Log(N$\rm _{H,l.o.s}$)=23.86$_{-0.11}^{+0.21}$. We also find that the obscuring torus is likely to be homogenous, since the column density responsible for the scattered emission (which can be treated as the torus average column density) is Log(N$\rm _{H,S}$)=24.14$_{-0.28}^{+0.31}$, in good agreement with the l.o.s. one.

We complete our spectral analysis using \borus, and even in this case the fit statistic is fully consistent with those of the two \myt\ fits ($\chi^2$/d.o.f.=117.8/122). The best fit parameters are also in excellent agreement with the \myt\ ones: the photon index is  $\Gamma$=1.56$_{-0.16}^{+0.20}$, the l.o.s. column density is Log(N$\rm _{H,l.o.s}$)=23.85$_{-0.10}^{+0.08}$ and the torus average column density for which we obtain the smallest $\chi^2$ is Log(N$\rm _{H,tor}$)=24.1. Since ESO 317-IG 030 is significantly more obscured than ESO 244-IG 030, we are also able to put some (albeit loose) constraints on the torus covering factor, $f_c$=0.58$_{-0.30}^{+0.42}$. We note that in \myt\ coupled the torus covering factor is also assumed to be $f_c$=0.50. The best fit model is shown in Figure \ref{fig:spectra}, central right panel.

In summary, the fit of the \nus\ and \xmm\ data depicts a scenario where ESO 317-IG 030 has a harder spectrum and is significantly less obscured than it was measured using the low-quality \xrt\ and \swi\ data. Based on our new measurements, ESO 317-IG 030 is reclassified as a Compton thin AGN, although the 90--95\% (depending on the model) upper boundary of the l.o.s. column density is above the Log(N$\rm _{H,l.o.s}$)=24 threshold.

\subsection{2MASX J10331570+5252182}\label{sec:J1033}
Before performing our spectral analysis, we check if the assumptions on the source counterpart and redshift re-association reported in Section \ref{sec:sample} are correct: to do so, we fit our data leaving $z$ as a free parameter. Thanks to the high spectral quality of the \xmm\ and \nus\ data, we are able to tightly constrain $z$: with all three models used in our analysis, we find a redshift value in excellent agreement with $z_{\rm spec}$=0.14036, i.e., the spectroscopic redshift measured for 2MASX J10331570+5252182. More in detail, when using \myt\ in its ``coupled'' configuration we obtain $z_{\rm MyTC}$=0.147$_{-0.012}^{+0.008}$, with \myt\  in ``decoupled'' mode we measure a redshift $z_{\rm MyTC}$=0.145$_{-0.007}^{+0.010}$, and with \borus\ we find $z_{\rm bor}$=0.147$_{-0.009}^{+0.008}$. We also point out that using the previous redshift value ($z$=0.0653) significantly affects the fit statistic, with a variation in $\chi^2$ $\Delta\chi^2$$\sim$40, regardless of the adopted model.
For these reasons, in the rest of the analysis we fix the source redshift to $z_{\rm spec}$=0.14036.

As for the previous two objects, we first fit the X-ray spectrum using ``\myt\ coupled'': the best fit statistic we obtain is good, being $\chi^2$/d.o.f.=128.8/114. We find that the photon index is pegged at the parameter lower boundary, $\Gamma$=1.4, i.e., that the source is particularly hard. Such a hard photon index may be unphysical \citep[the typical AGN photon index is $\Gamma$=1.7--1.9, see, e.g.,][]{marchesi16c} and could imply that the l.o.s. column density measurement we obtain is underestimated.  While in ESO 244-IG 030 and ESO 317-IG 030 the fit is insensitive to $\theta_{\rm obs}$, which we thus fixed to $\theta_{\rm obs}$=90$\degree$, in 2MASX J10331570+5252182 the fit is significantly improved by allowing the angle between the observer and the torus axis, $\theta_{\rm obs}$, to vary. Particularly, the best-fit solution is $\theta_{\rm obs}$=61.0$_{-0.7}^{+4.0}$$\degree$, i.e., the source would be observed at the edge of the torus, whose opening angle is $\theta_{\rm OA}$=60$\degree$: this is a scenario commonly observed when using ``\myt\ coupled'' \citep[see, e.g.,][]{zhao19b,zhao19a}, but is also physically unlikely, since the chance of observing the accreting SMBH exactly at the edge of the obscuring torus is small. The equatorial column density we measure is N$\rm _{H,eq}$= 2.29$_{-0.93}^{+1.07}$ $\times$ 10$^{24}$\,cm$^{-2}$,  and the corresponding l.o.s. column density is therefore N$_{\rm H,l.o.s.}$ = N$_{\rm H,eq}$ [1 - 4 cos$^2\theta_{\rm obs}$]$^{1/2}$=4.89$_{-2.27}^{+2.62}$ $\times$ 10$^{23}$\,cm$^{-2}$.

\begin{figure*}
\begin{minipage}[b]{.5\textwidth}
  \centering
  \includegraphics[width=0.78\textwidth,angle=-90]{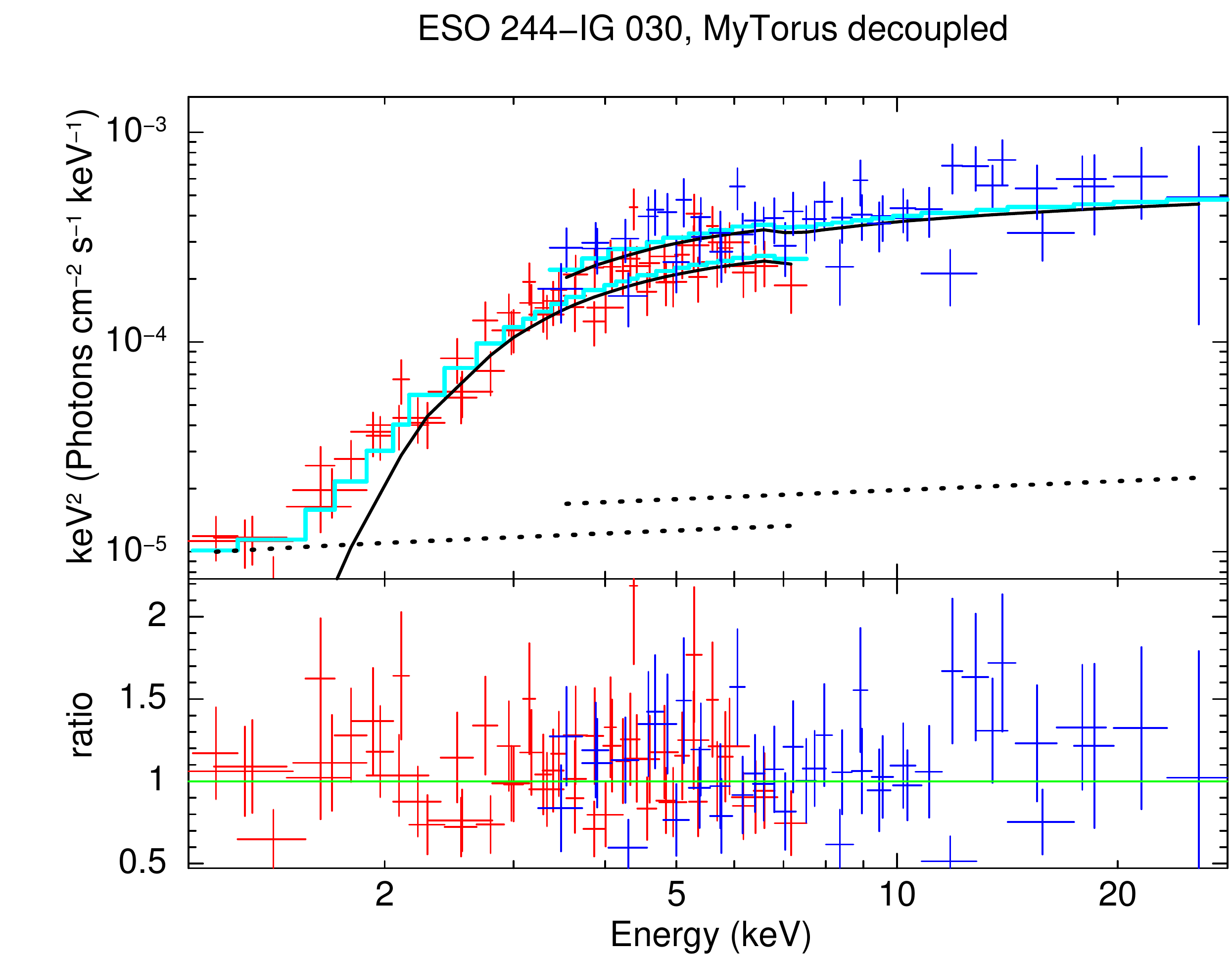}
  \end{minipage}
\begin{minipage}[b]{.5\textwidth}
  \centering
  \includegraphics[width=0.78\textwidth,angle=-90]{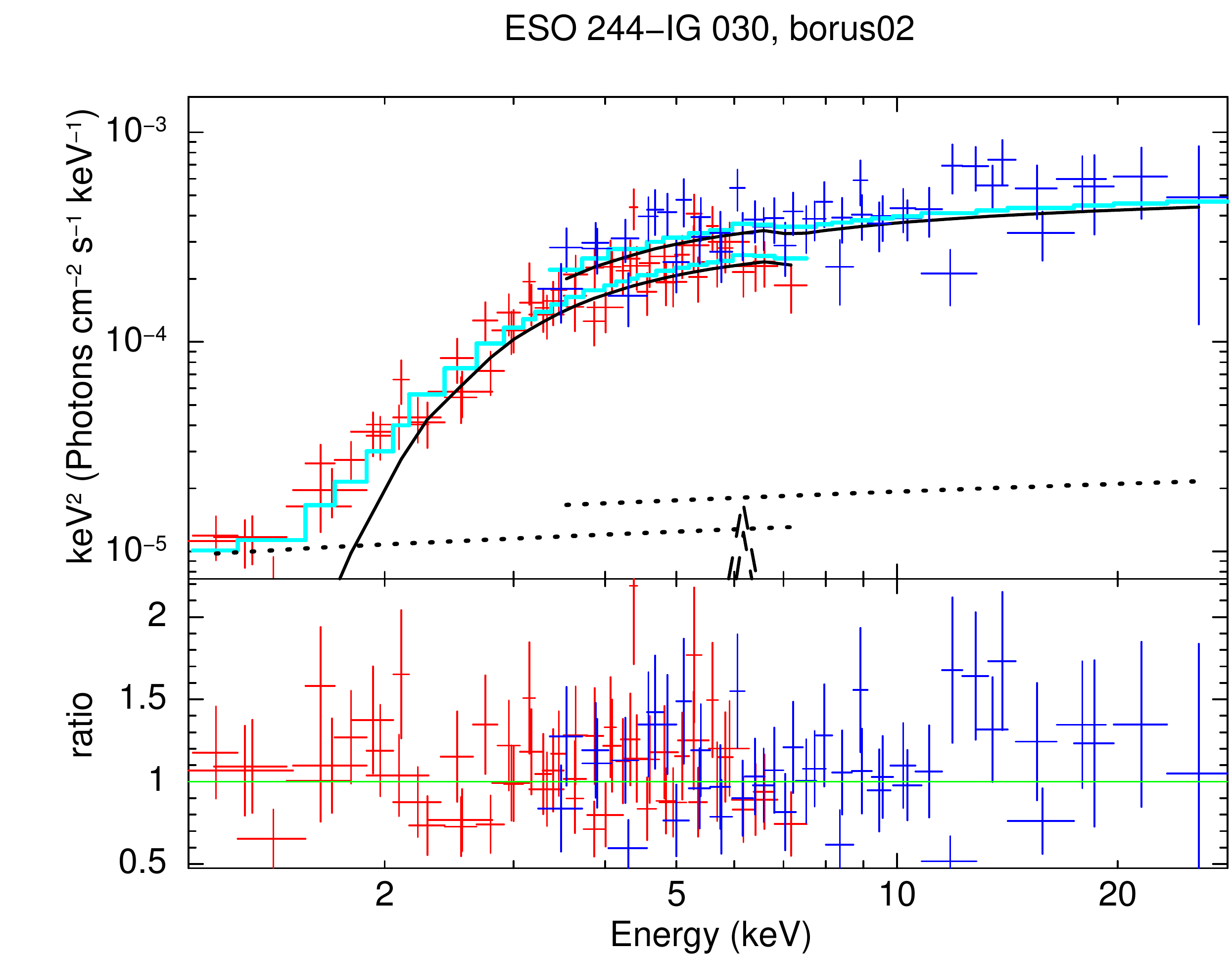}
  \end{minipage}
  \begin{minipage}[b]{.5\textwidth}
  \centering
  \includegraphics[width=0.78\textwidth,angle=-90]{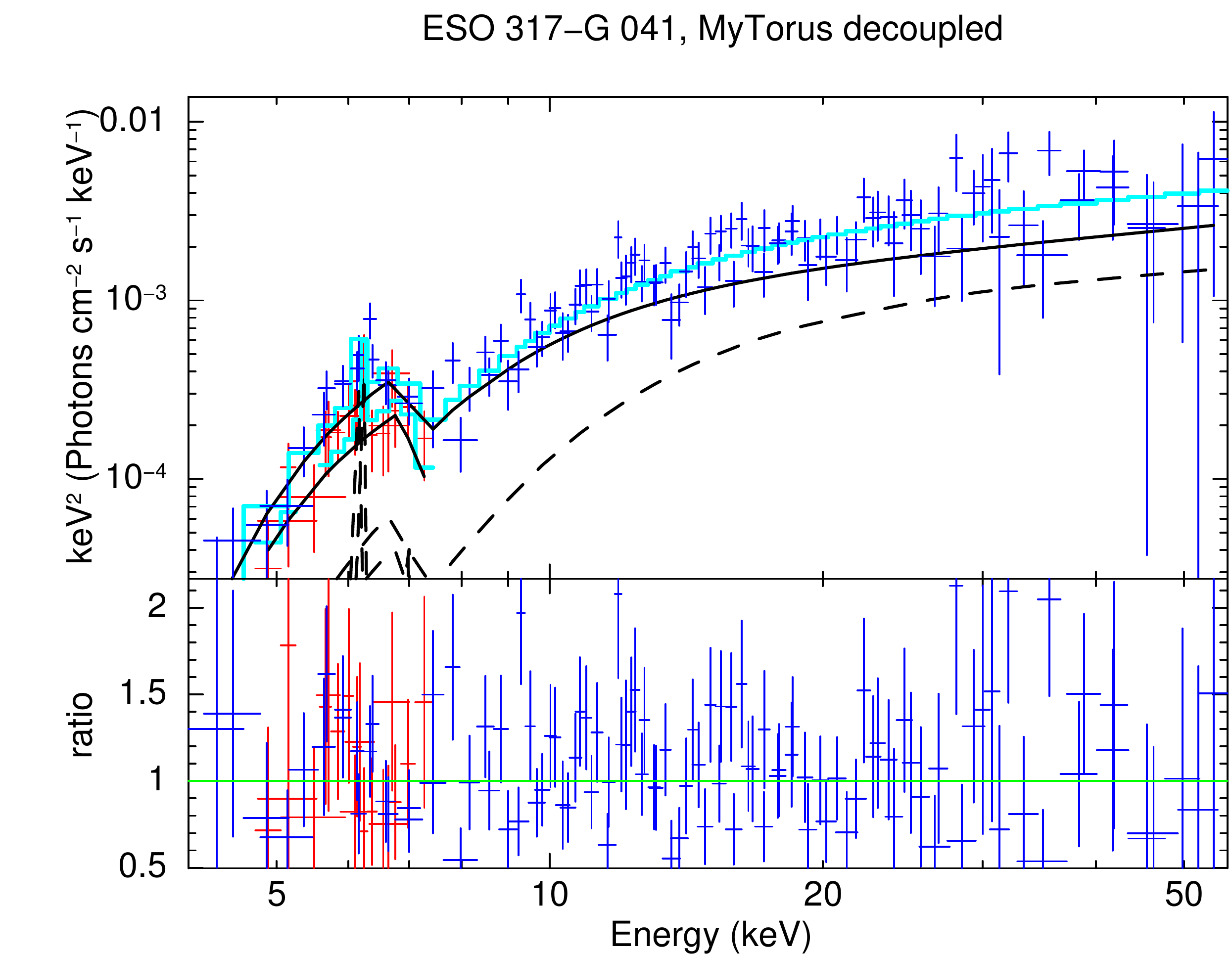}
  \end{minipage}
  \begin{minipage}[b]{.5\textwidth}
  \centering
  \includegraphics[width=0.78\textwidth,angle=-90]{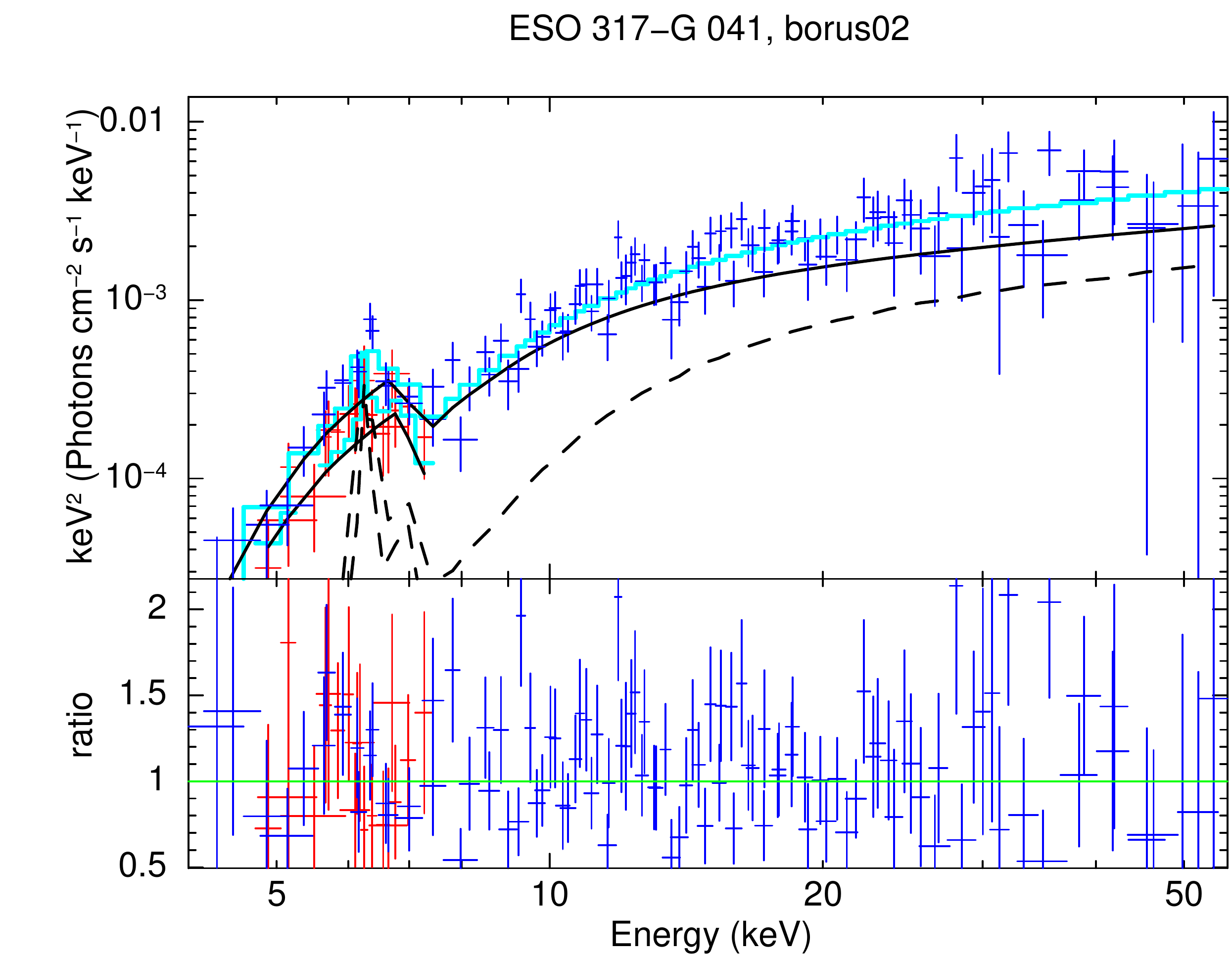}
  \end{minipage}
  \begin{minipage}[b]{.5\textwidth}
  \centering
  \includegraphics[width=0.78\textwidth,angle=-90]{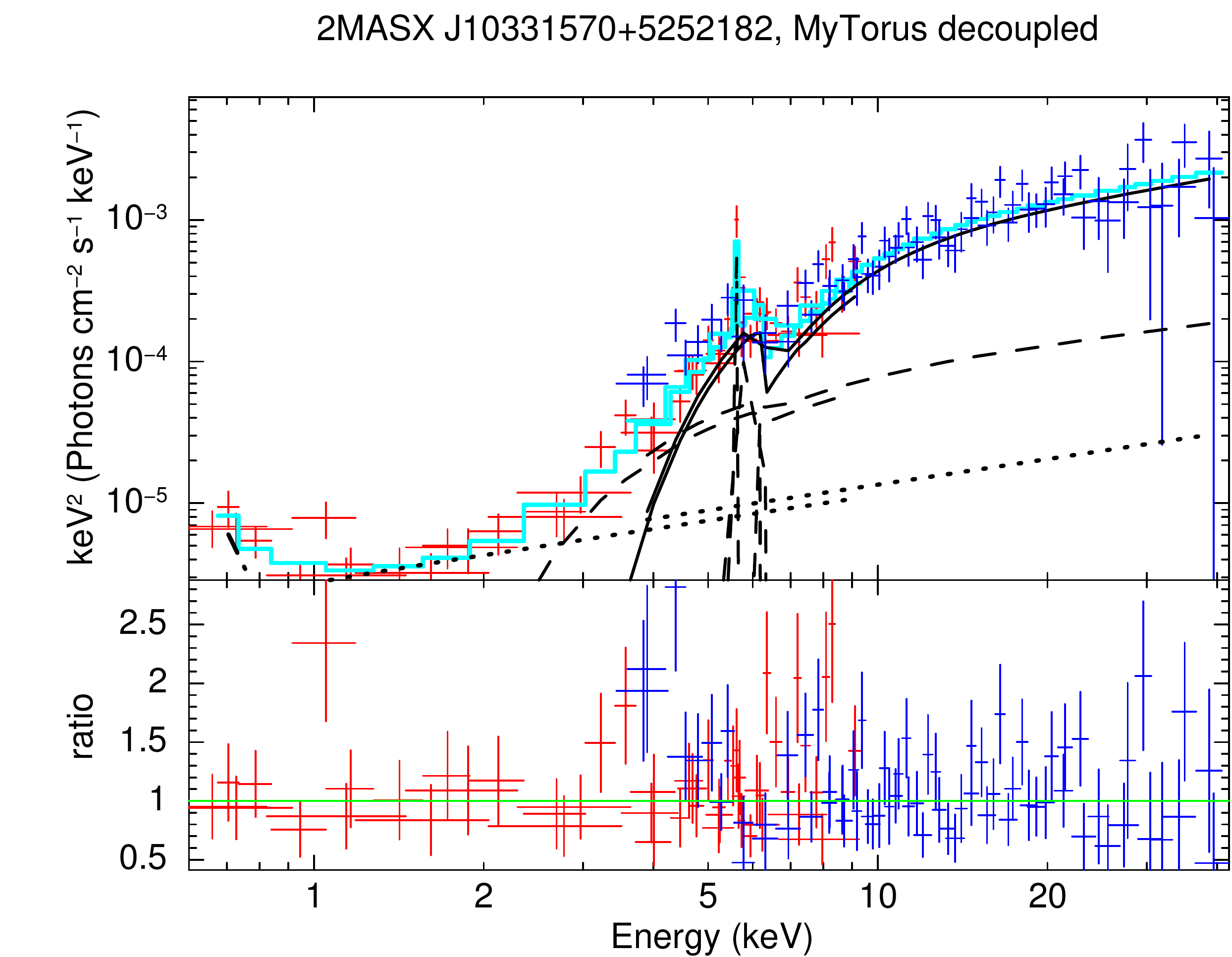}
  \end{minipage}
  \begin{minipage}[b]{.5\textwidth}
  \centering
  \includegraphics[width=0.78\textwidth,angle=-90]{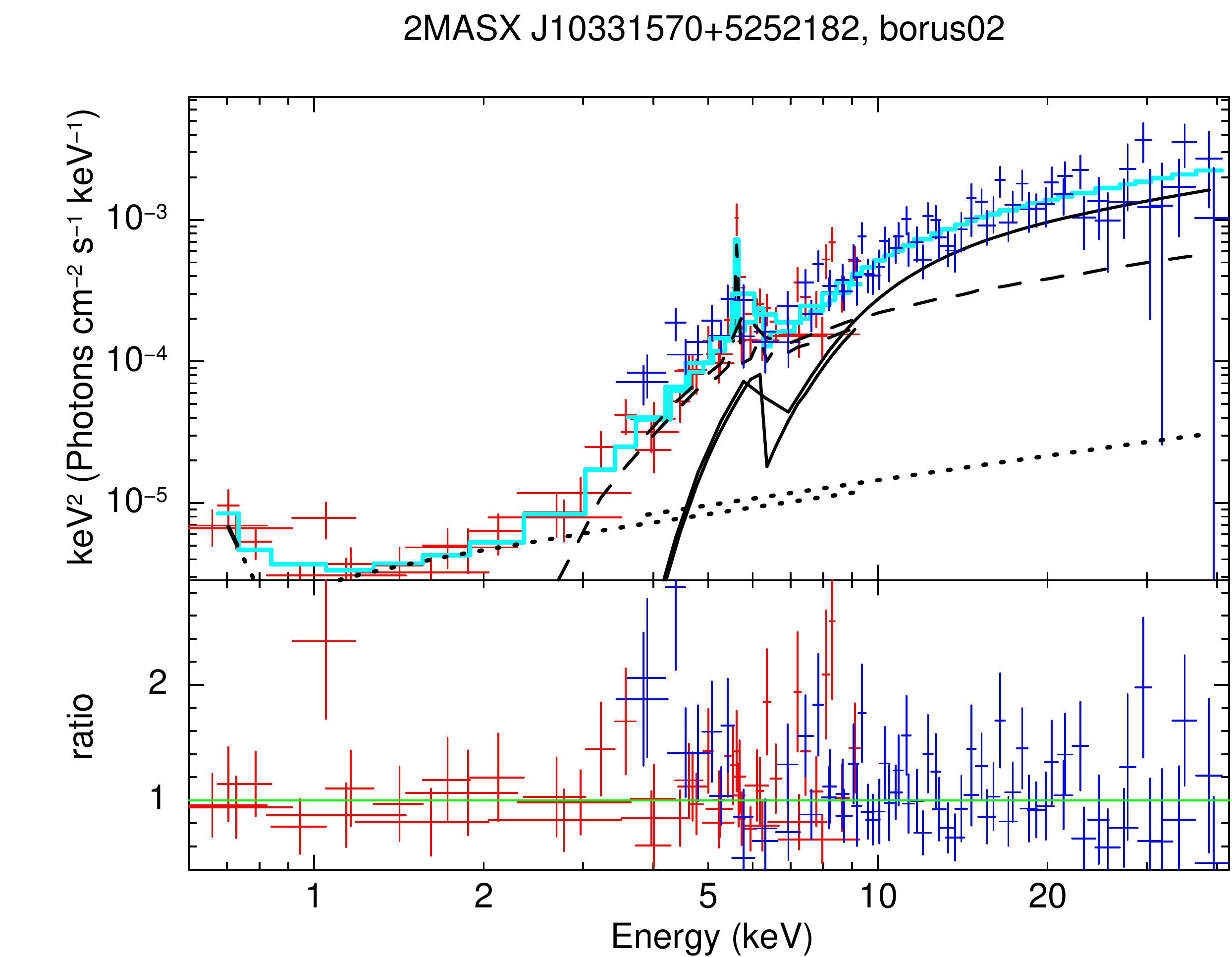}
  \end{minipage}
  \caption{\small Background-subtracted spectra (top panel) and data-to-model ratio (bottom) of the three sources in this work. \xmm\ data are plotted in red, \nustar\ data in blue. In the left column we report the best-fit models obtained using \myt\ in ``decoupled'' configuration, while in the right column we show the \borus\ best-fit models. The best-fitting model is plotted as a cyan solid line, the main emission component is plotted as a black solid line, the reprocessed component is plotted as a black dashed line and the main power law component scattered, rather than absorbed, by the torus is plotted as a black dotted line. Finally, the phenomenological \texttt{mekal} component used to model the excess below 1\,keV is plotted as a dash-dotted black line.}\label{fig:spectra}
\end{figure*}

When fitting the 2MASX J10331570+5252182 0.6--50\,keV spectrum with ``\myt\ decoupled'' we again find that the photon index is pegged to the parameter lower boundary, $\Gamma$=1.4. The l.o.s. column density is slightly higher, although consistent within the 90\,\% confidence uncertainties, than the one measured using ``\myt\ coupled'', being Log(N$\rm _{H,l.o.s.,MyTD}$)=23.89$_{-0.07}^{+0.08}$. Interestingly, the column density linked to the reprocessed emission is found to be significantly smaller (Log(N$\rm _{H,S}$)=23.05$_{-0.35}^{+0.24}$), potentially hinting to a scenario where the accreting SMBH is observed through an overdensity in a patchy torus. We show the best-fit ``\myt\ decoupled'' model in Figure \ref{fig:spectra}, bottom left panel.

Finally, we find that using \borus\ the fit statistic is significantly improved when allowing the metallicity of the \borus\ model to vary with respect to the Solar value: the best fit metallicity is Z/Z$_\odot$=0.35$_{-0.15}^{+0.16}$. The other parameters are in good agreement with those measured using ``\myt\ decoupled'': the photon index is pegged at $\Gamma$=1.4 and the l.o.s column density is just above the CT threshold (Log(N$\rm _{H,l.o.s.}$)=24.07$_{-0.14}^{+0.12}$). The average torus column density is also in good agreement with the value obtained using ``\myt\ decoupled'', being Log(N$\rm _{H,tor}$)=23.4. All three models have similar fit statistic, but we favor the ``\myt\ decoupled'' and \borus\ solutions given the low physical likelihood of the scenario depicted by ``\myt\ coupled''. The best fit model is shown in Figure \ref{fig:spectra}, bottom right panel.

Interestingly, we find a second possible best fit model for 2MASX J10331570+5252182 using \borus: in this second scenario (whose fit statistic is $\chi^2$/d.o.f.=124.9/144, with a difference in $\chi^2$ $\Delta$$\chi^2$=0.3 with respect to the first solution), the accreting SMBH is buried in a heavily CT torus with Log(N$\rm _{H,tor}$)=25 and Log(N$\rm _{H,l.o.s.}$)=25.05$_{-0.13}^{+0.95}$. The best fit photon index is  $\Gamma$=1.90$_{-0.14}^{+0.19}$ and the torus covering factor is pegged at the parameter lower boundary, $f_c$=0.1. While these parameters are all physically plausible, the intrinsic luminosity associated to the source is extremely high, being Log(L$_{2-10}$)$>$46, i.e., almost an order of magnitude higher than the luminosities observed in luminous quasars. For this reason, we rule out this high-luminosity solution, which nonetheless confirms that 2MASX J10331570+5252182 must be heavily obscured.

We also point out that, although in our best-fit solution the l.o.s. column density is below the CT threshold, the fact that the photon index is pegged at $\Gamma$=1.4 suggests that we might be underestimating N$\rm _{H,l.o.s.}$, and 2MASX J10331570+5252182 could be a CT-AGN. As a test, we refit our spectrum fixing the photon index to a typical AGN value, $\Gamma$=1.8. In all three models,  the variation in $\chi^2$ with respect to the best fit model with $\Gamma$=1.4 is $\Delta$$\chi^2$$\sim$11.5 and the reduced $\chi^2$ is $\chi^2_\nu$$\sim$1.19: the $\Gamma$=1.8 is therefore statistically plausible, although less likely than the $\Gamma$=1.4 one. 

When fixing the photon index to $\Gamma$=1.8, we find a 30--35\,\% increase in N$\rm _{H,l.o.s.}$ and a corresponding increase in the main power law normalization by a factor of $\sim$4, regardless of the adopted model. For all the other parameters the best-fit values are consistent, within the uncertainties, with those obtained using $\Gamma$=1.4. Using ``\myt\ decoupled'' and ``\borus'', the increase in l.o.s. column density shifts the parameter best-fit value above the CT threshold, Log(N$\rm _{H,l.o.s.}$)$>$24.

In conclusion, all the models used in our analysis are statistically acceptable and they all depict a heavily obscured, likely CT scenario for 2MASX J10331570+5252182. This makes this source fairly unique among those in the 100-month BAT catalog, since \swi\ is biased against obscured AGN: for example, \citet{burlon11} reported that in the 36-month BAT catalog no CT-AGNs were detected at redshifts $z$$>$0.04; similarly, only 3 out of 55 candidate CT-AGNs in the 70-month BAT catalog have redshift $z$$>$0.1 \citep{ricci15}, one of which (2MASXJ03561995$-$6251391) was reclassified as Compton thin in M18 and M19. We further discuss the potential uniqueness of 2MASX J10331570+5252182 in Section \ref{sec:lumin}.

\begingroup
\renewcommand*{\arraystretch}{1.5}
\begin{table*}
\centering
\scalebox{0.95}{
\vspace{.1cm}
  \begin{tabular}{cccc}
       \hline
       \hline       
       Source & ESO 244-IG 030 & ESO 317-G 041 & 2MASX J10331570+5252182\\
       \hline
       %
       \multicolumn{4}{c}{\myt\ coupled}\\
       $\chi^2$/d.o.f.& 104.4/97 & 118.4/122 & 128.8/114 \\
       $C_{Ins}$& 1.41$_{-0.16}^{+0.17}$ & 1.61$_{-0.27}^{+0.35}$ & 1.22$_{-0.09}^{+0.13}$ \\
       $z$ & 0.0256$^f$ & 0.0193$^f$ & 0.14036$^f$ \\
       $\Gamma$ & 1.86$_{-0.16}^{+0.17}$ & 1.47$_{-0.07l}^{+0.25}$ & 1.40$^f$ \\
       norm 10$^{-4}$ & 2.34$_{-0.62}^{+0.87}$ & 4.39$_{-1.44}^{+6.42}$ & 1.73$_{-0.51}^{+0.80}$ \\
       $\theta\rm _{obs}$& 90$^f$ & 90$^f$ & 61.0$_{-0.7}^{+4.0}$ \\
       Log(N$\rm _{H,eq}$) & 22.75$_{-0.08}^{+0.08}$ & 23.88$_{-0.05}^{+0.08}$ &  24.30$_{-0.27}^{+0.16}$ \\
       Log(N$\rm _{H,l.o.s.}$) & 22.75$_{-0.08}^{+0.08}$ & 23.88$_{-0.05}^{+0.08}$ & 23.69$_{-0.27}^{+0.16}$ \\
       A$_S$ & 1.00$^f$ & 0.93$_{-0.59}^{+0.89}$ & 1.66$_{-0.78}^{+1.03}$  \\
       $f_s$ 10$^{-2}$& 4.9$_{-1.3}^{+1.8}$ & -- & 1.7$_{-0.3}^{+0.5}$ \\
       $kT$ & -- & -- & 0.28$_{-0.10}^{+0.23}$ \\
       F$_{2-10}$ & 4.63$_{-0.45}^{+0.27}$ & 2.47$_{-1.83}^{+0.28}$ & 2.92$_{-0.23}^{+12.57}$ \\
       F$_{15-55}$ & 10.12$_{-2.41}^{+1.72}$ & 62.58$_{-46.76}^{+1.46}$  & 38.80$_{-6.70}^{+15.88}$ \\
       Log(L$_{2-10}$) & 42.00$_{-0.10}^{+0.08}$ & 42.42$_{-0.34}^{+0.19}$ & 43.74$_{-0.09}^{+0.23}$ \\
       Log(L$_{15-55}$) & 42.02$_{-0.51}^{+0.34}$ & 42.68$_{-0.43}^{+0.29}$ & 44.43$_{-0.80}^{+0.27}$ \\
       \hline
       %
       \multicolumn{4}{c}{\myt\ decoupled ``edge-on''}\\
       $\chi^2$/d.o.f.& 104.2/97 & 117.9/121 & 126.6/114 \\
       $C_{Ins}$ & 1.41$_{-0.16}^{+0.17}$ & 1.64$_{-0.28}^{+0.36}$ & 1.18$_{-0.16}^{+0.18}$ \\
       $\Gamma$& 1.86$_{-0.16}^{+0.17}$ & 1.60$_{-0.20l}^{+0.34}$ & 1.40$^f$ \\
       norm 10$^{-4}$ & 2.13$_{-0.57}^{+0.82}$ & 5.39$_{-2.98}^{+11.68}$ & 3.19$_{-0.75}^{+0.99}$ \\       
       A$_S$ & 1.00$^f$ & 1.75$_{-1.45}^{+3.15}$ & 1.99$_{-0.70}^{+0.85}$ \\
       Log(N$\rm _{H,l.o.s.}$) & 22.76$_{-0.08}^{+0.07}$ & 23.86$_{-0.11}^{+0.21}$ & 23.89$_{-0.07}^{+0.08}$ \\
       Log(N$\rm _{H,S}$) & -- & 24.14$_{-0.28}^{+0.31}$ & 23.05$_{-0.35}^{+0.24}$ \\
       $f_s$ 10$^{-2}$ & 4.7$_{-1.3}^{+1.8}$ & -- & 0.9$_{-0.3}^{+0.4}$ \\
       F$_{2-10}$ & 4.71$_{-0.84}^{+0.66}$ & 2.44$_{-1.82}^{+0.27}$ & 3.00$_{-0.36}^{+0.24}$ \\
       F$_{15-55}$ & 10.14$_{-2.89}^{+2.69}$ & 61.47$_{-46.93}^{+2.78}$ & 37.25$_{-4.14}^{+2.96}$ \\
       Log(L$_{2-10}$) & 42.00$_{-0.09}^{+0.07}$ & 42.40$_{-0.49}^{+0.24}$ & 44.01$_{-0.09}^{+0.07}$ \\
       Log(L$_{15-55}$) & 42.01$_{-0.53}^{+0.34}$ & 42.64$_{-0.52}^{+0.37}$ & 44.74$_{-0.71}^{+0.26}$ \\
       Log(L$_{bol}$) & 43.00$_{-0.10}^{+0.08}$ & 43.38$_{-0.49}^{+0.25}$ & 45.16$_{-0.12}^{+0.10}$ \\
              \hline
       %
       \multicolumn{4}{c}{\borus}\\
       $\chi^2$/d.o.f.& 104.2/97 & 117.8/122 & 124.6/114 \\
       $C_{Ins}$ & 1.41$_{-0.15}^{+0.17}$ & 1.64$_{-0.27}^{+0.37}$ & 1.16$_{-0.13}^{+0.13}$ \\ 
       $\Gamma$& 1.85$_{-0.17}^{+0.17}$ & 1.56$_{-0.16l}^{+0.20}$ & 1.40$^f$ \\ 
       norm 10$^{-4}$ & 2.10$_{-0.58}^{+0.85}$ & 4.87$_{-1.96}^{+5.06}$ & 3.98$_{-0.86}^{+1.28}$ \\ 
       Log(N$\rm _{H,l.o.s.}$) & 22.77$_{-0.09}^{+0.07}$ & 23.85$_{-0.10}^{+0.08}$ & 24.07$_{-0.14}^{+0.12}$ \\ 
       Log(N$\rm _{H,tor}$) & 22.2$^f$ & 24.1$^f$ & 23.4$^f$ \\ 
       Z/Z$_\odot$ & 1.00$^f$  & 1.00$^f$  & 0.35$_{-0.15}^{+0.16}$  \\ 
       $f_c$ & 1.00$^f$ & 0.58$_{-0.30}^{+0.42}$ & 0.90$_{-0.24}^{+0.10u}$ \\ 
       $f_s$ 10$^{-2}$& 4.7$_{-1.3}^{+1.7}$ & -- & 0.8$_{-0.3}^{+0.3}$ \\ 
       F$_{2-10}$ & 4.71$_{-0.46}^{+0.25}$ & 2.44$_{-1.08}^{+0.27}$ &  3.01$_{-0.78}^{+0.32}$ \\ 
       F$_{15-55}$ & 9.79$_{-2.16}^{+1.54}$ & 61.73$_{-29.01}^{+2.50}$ & 38.61$_{-8.66}^{+2.82}$ \\ 
       Log(L$_{2-10}$) & 42.00$_{-0.09}^{+0.07}$ & 42.36$_{-0.39}^{+0.34}$ & 44.11$_{-0.09}^{+0.08}$ \\
       Log(L$_{15-55}$) & 42.01$_{-0.49}^{+0.35}$ & 42.60$_{-0.45}^{+0.32}$ & 44.86$_{-1.11}^{+0.28}$ \\
       \hline
	\hline
	\vspace{0.02cm}
\end{tabular}}
	\caption{\small \raggedright Best-fits results for the joint \nus--\xmm\ spectral fitting of the three objects in our sample. $C_{Ins}$ = $C_{NuS/XMM}$ is the cross calibration between \NuSTAR\ and \XMM, $z$ is the source redshift, $\Gamma$ is the power law photon index, norm is the main power law normalization at 1\,keV, in units of photons s$^{-1}$ cm$^{-2}$ keV$^{-1}$, N$\rm _{H,eq}$ is the equatorial column density, in units of cm$^{-2}$,  $\theta\rm _{obs}$ is the inclination angle between the observer line of sight and the torus axis (in degrees), A$_S$ is the intensity of the reprocessed component, $f_s$ is the fraction of emission scattered, rather than absorbed by the obscuring torus, $kT$ is the temperature (in keV) of the phenomenological \texttt{mekal} component used to fit the excess at $<$1\,keV. N$\rm _{H,l.o.s.}$ is the line of sight column density, while N$\rm _{H,S}$ and N$\rm _{H,tor}$ are the average torus column densities as measured using \myt\ decoupled and \borus, respectively, and $f_c$ is the torus covering factor, $f_c$ = cos($\theta_{\rm tor}$); Z is the metallicity of the \borus\ component. Finally, F$_{2-10}$ and F$_{15-55}$ are the observed fluxes (in units of $10^{-13}$ erg cm$^{-2}$ s$^{-1}$) in the 2--10\,keV and 15--55\,keV bands, while L$_{2-10}$ and L$_{15-55}$ are the intrinsic luminosities (in units of erg s$^{-1}$) in the same bands, and L$_{bol}$ is the bolometric luminosity (in units of erg s$^{-1}$) derived using L$_{2-10}$ and the bolometric correction for Type 2 AGN derived by \citet{lusso12}.}
\label{tab:results}
\end{table*}
\endgroup

\section{Discussion and conclusions}\label{sec:concl}

\subsection{The key role of \nus\ in the characterization of heavily obscured AGN}
In M18 and M19 we showed how the addition of \nus\ significantly improves the results of the X-ray spectral analysis of heavily obscured AGNs. To check if this evidence is confirmed also in our new sample of object, we re-fit the \xmm\ spectra alone.

ESO 244--IG 030 would have been found to be a non-CT AGN even using the \xmm\ data alone. However, while the best fit parameters measured with \xmm\ are in good agreement with those derived using both \xmm\ and \nus\ ($\Gamma_{\rm XMM}$=1.89$_{-0.39}^{+0.44}$; Log(N$_{\rm H,l.o.s.,XMM}$)=22.78-0.13+0.11), the uncertainties on the parameters are significantly higher: the $\Gamma$ uncertainty increases by a factor $\sim$2, and we measure a $\sim$50\,\% increase in the N$\rm _{H,l.o.s.}$ uncertainty. Finally, as expected, the source covering factor is unconstrained.

ESO 317--G 041 is the source in our sample where the lack of \nus\ data would affect the fit reliability the most, since the source \xmm\ count statistic in the 3--10\,keV band is fairly poor ($\sim$300 counts overall). The photon index is pegged to the model lower boundary ($\Gamma$=1.4), while the uncertainties on N$\rm _{H,l.o.s.}$ are a factor $\sim$2.5 larger than those computed using the joint \nus\ and \xmm\ spectra. Finally, no constrain can be put on both the torus average column density and on the covering factor based on the \xmm\ data only.

Finally, for 2MASX J10331570+5252182 the ``\xmm-only'' best-fit results are once again in good agreement with the joint \xmm\ and \nus\ ones ($\Gamma_{\rm XMM}$=1.4 and Log(N$_{\rm H,l.o.s.,XMM}$)=24.15$_{-0.19}^{+0.20}$), but the uncertainties on N$_{\rm H,l.o.s.,XMM}$ are $\sim$40\,\% larger. Furthermore, without the \nus\ data $f_c$ and  N$\rm _{H,tor}$ are fully unconstrained.

Summarizing, while the \xmm\ data alone would have allowed one to reliably measure the main spectral parameters ($\Gamma$ and N$_{\rm H,l.o.s}$), the parameter uncertainties would have been significantly larger, thus limiting the significance of the fit results. Furthermore, the lack of \nus\ would have prevented one to measure other important parameters, particularly the torus covering factor and average column density.

\subsection{Measurements of the Iron K$\alpha$ line equivalent width}\label{sec:EW}
Historically, candidate heavily obscured AGNs have been selected in the X-rays based on the presence of a prominent Iron K$\alpha$ line at 6.4\,keV, having equivalent width EW$\gtrsim$1\,keV: however, the low spectral quality of the \xrt\ data for the three sources in our sample did not allow us to constrain the line EW, a goal now achievable using the \nus\ and \xmm\ data. To do so, since neither \myt\ nor \borus\ allow one to use the task \texttt{eqwidth} in \texttt{XSPEC}, we follow the approach described in \citet{yaqoob15} to compute the Iron line EW using \myt. We thus first compute the continuum flux at 6.4\,keV, without including the contribution of the emission line. Then, we measure the flux of the Iron line in the energy range E = [0.95\,E$_{\rm K\alpha}$--1.05\,E$_{\rm K\alpha}$] (i.e., between 6.08 and 6.72\,keV, rest frame). The line EW is then obtained by multiplying by (1+$z$) the ratio between the line flux and the continuum flux at 6.4\,keV.

As reported in Section \ref{sec:eso244}, ESO 244-IG 030 is only moderately obscured, having Log(N$\rm _{H,l.o.s.}$)$\leq$22.8: consequently, this source does not have a prominent Iron line, and in fact no clear emission feature is observed at E$\sim$6.4\,keV. 
The upper limit on the line equivalent width is EW$_{\rm K\alpha,MyTC}$$\leq$0.05\,keV and EW$_{\rm K\alpha,MyTD}$$\leq$0.09\,keV using either ``\myt\ coupled'' or ``decoupled'', respectively.

ESO 317-G 041 is significantly more obscured than ESO 244-IG 030, having Log(N$\rm _{H,l.o.s.}$)$\sim$23.85, and an emission feature is visible in its spectrum at $\sim$6.4\,keV. The equivalent width of the line is however slightly smaller than the values typically observed in CT-AGNs, being EW$_{\rm K\alpha,MyTC}$=0.21$_{-0.19}^{+0.03}$\,keV using ``\myt\ coupled and EW$_{\rm K\alpha,MyTC}$$<$0.27\,keV with ``\myt\ decoupled''.

As can be seen, the Iron K$\alpha$EW measurements further confirm that both ESO 244-IG 030 and ESO 317-G 041 have significantly lower l.o.s. column density than it was expected based on the  \xrt\ and \swi\ spectral fitting.

Finally, 2MASX J10331570+5252182 is the source in our sample where the Iron line is most clearly visible in the spectrum and, consequently, the source where the line EW is the largest: with ``\myt\ coupled'' (``decoupled'') we measure EW$_{\rm K\alpha,MyTC}$=0.34$\pm$0.07\,keV (EW$_{\rm K\alpha,MyTD}$=0.30$_{-0.12}^{+0.11}$\,keV). We also compute EW using the best fit models obtained fixing the photon index to $\Gamma$=1.8, and the results are in good agreement with those obtained using the $\Gamma$=1.4 fits, being EW$_{\rm K\alpha,MyTC,\Gamma=1.8}$=0.39$_{-0.18}^{+0.09}$\,keV and EW$_{\rm K\alpha,MyTD,\Gamma=1.8}$=0.21$_{-0.05}^{+0.20}$\,keV.

\subsection{Intrinsic X-ray and bolometric luminosity, and the SMBH mass}\label{sec:lumin}
In Table \ref{tab:results} we report the intrinsic, absorption-corrected luminosities of the three sources in our sample in the 2--10\,keV and 15--55\,keV bands. For all sources, all the three models used in our analysis produce luminosities values that are in agreement within the 90\,\% uncertainties. 

Both ESO 244-IG 030 and ESO 317-G 041 are low/moderate luminosity AGN: the first has L$_{\rm X}$$\sim$10$^{42}$ erg s$^{-1}$ in both the 2--10\,keV and in the 15--55\,keV band, while the latter has L$_{\rm 2-10}$$\sim$2.5\,$\times$\,10$^{42}$\,erg s$^{-1}$ and L$_{\rm 15-55}$$\sim$4.5\,$\times$\,10$^{42}$\,erg s$^{-1}$. 
While 2MASX J10331570+5252182 is far more distant than the other two sources in our sample ($z$=0.14036 versus $z$$\sim$0.019--0.026), all three objects have similar observed fluxes: consequently, 2MASX J10331570+5252182 is significantly more luminous than the other two, and is in fact a candidate Compton thick quasar, having L$_{\rm 2-10}$$\sim$10$^{44}$\,erg s$^{-1}$ and L$_{\rm 15-55}$$\sim$7\,$\times$\,10$^{44}$\,erg s$^{-1}$. As we mentioned in Section \ref{sec:J1033}, MASX J10331570+5252182 is one of the very few candidate CT-AGN detected by \swi\ at $z$$>$0.1: its quasar-like luminosity is also uncommon in heavily obscured AGN detected by \swi, since among the 35 100-month BAT-selected, candidate CT-AGNs studied in M18, M19, only two have Log(L$_{\rm 2-10}$)$>$44 (namely, MASX~J03561995--6251391 and MRK 3).

The intrinsic 2--10\,keV luminosities we derived from our spectral fitting can be used to estimate the AGN bolometric luminosity (i.e., the total AGN emission integrated over the whole electromagnetic spectrum), using one of the several bolometric corrections reported in the literature \citep[see, e.g.,][]{elvis94,marconi04,lusso12,brightman17}: in this work, we use the bolometric correction for Type 2 AGN computed by \citet[][we use their ``spectro+photo'' sample parameters]{lusso12}, and the 2--10\,keV luminosities derived using the ``\myt\ decoupled'' model. The bolometric luminosities we obtain are L$\rm _{bol}$=1.00$_{-0.21}^{+0.20}$\,$\times$\,10$^{43}$\,erg\,s$^{-1}$ for ESO 244-IG 030, L$\rm _{bol}$=2.40$_{-1.62}^{+1.87}$\,$\times$\,10$^{43}$\,erg\,s$^{-1}$ for ESO 317-G 041 and L$\rm _{bol}$=1.66$_{-0.40}^{+0.43}$\,$\times$\,10$^{45}$\,erg\,s$^{-1}$ for 2MASX J10331570+5252182. 

Finally, we use the bolometric luminosities derived using the \citet{lusso12} corrections to get a rough estimate of the masses of the accreting SMBHs powering the three AGNs in our sample: to do so, we assume an Eddington ratio $\lambda_{\rm Edd}$ = $L_{\rm bol}/L_{\rm Edd}$=0.1 \citep[a typical value for AGNs in the nearby Universe, see, e.g.,][]{marconi04}. In this equation, $L_{\rm Edd}$ is the Eddington luminosity, defined as 

\begin{equation}
L_{\rm Edd} = \frac{4\pi\,GM_{\rm BH}m_pc}{\sigma_T}, 
\end{equation}

where M$_{\rm BH}$ is the SMBH mass and m$_p$ is the mass of proton. The black hole mass is therefore 

\begin{equation}
M_{\rm BH} = \frac{L_{\rm bol}\,\sigma_T}{4\pi\,Gm_pc\lambda_{\rm Edd}}. 
\end{equation}

For the three objects in our sample, we estimate the following BH masses, in units of M$_\odot$: Log(M$_{\rm BH}$)=5.90$_{-0.10}^{+0.08}$ for ESO 244-IG 030; Log(M$_{\rm BH}$)=6.28$_{-0.49}^{+0.25}$ for ESO 317-G 041; and Log(M$_{\rm BH}$)=8.12$_{-0.12}^{+0.10}$ for 2MASX J10331570+5252182. The relatively low black holes masses we derive for ESO 244-IG 030 and ESO 317-G 041, while not physically implausible \citep[BH masses in AGN usually vary in the range log\,(M$\rm_{BH}$/M${_\sun}$) $\sim$6.0--9.8; see, e.g.,][]{Woo02}, may indicate that the Eddington ratio in these sources is lower than $\lambda_{\rm Edd}$=0.1, i.e., the objects are undergoing a less efficient accretion phase.

\subsection{Covering factor}\label{sec:cf}
As mentioned in the previous sections, the \borus\ model allows one to constrain the geometry of the obscuring material surrounding the accreting SMBH, using as a free parameter the torus covering factor, $f_c$. We were not able to put any significant constraint on $f_c$ in ESO 244-IG 030, since the source is only moderately obscured and the contribution to the total emission of the reprocessed component modeled by \borus\ is negligible. We are instead able to measure the torus covering factor for the other two sources in our sample: ESO 317-G 041 has  $f_c$= 0.58$_{-0.30}^{+0.42}$, close to the value adopted by \myt\ ($f_c$=0.5) and predicted by the so-called ``unified model'' for AGNs
2MASX J10331570+5252182, instead, favors a high covering factor solution, $f_c$=0.90$_{-0.24}^{+0.10u}$. 

\begin{figure*}
\begin{minipage}[b]{.33\textwidth}
  \centering
  \includegraphics[width=1.06\textwidth]{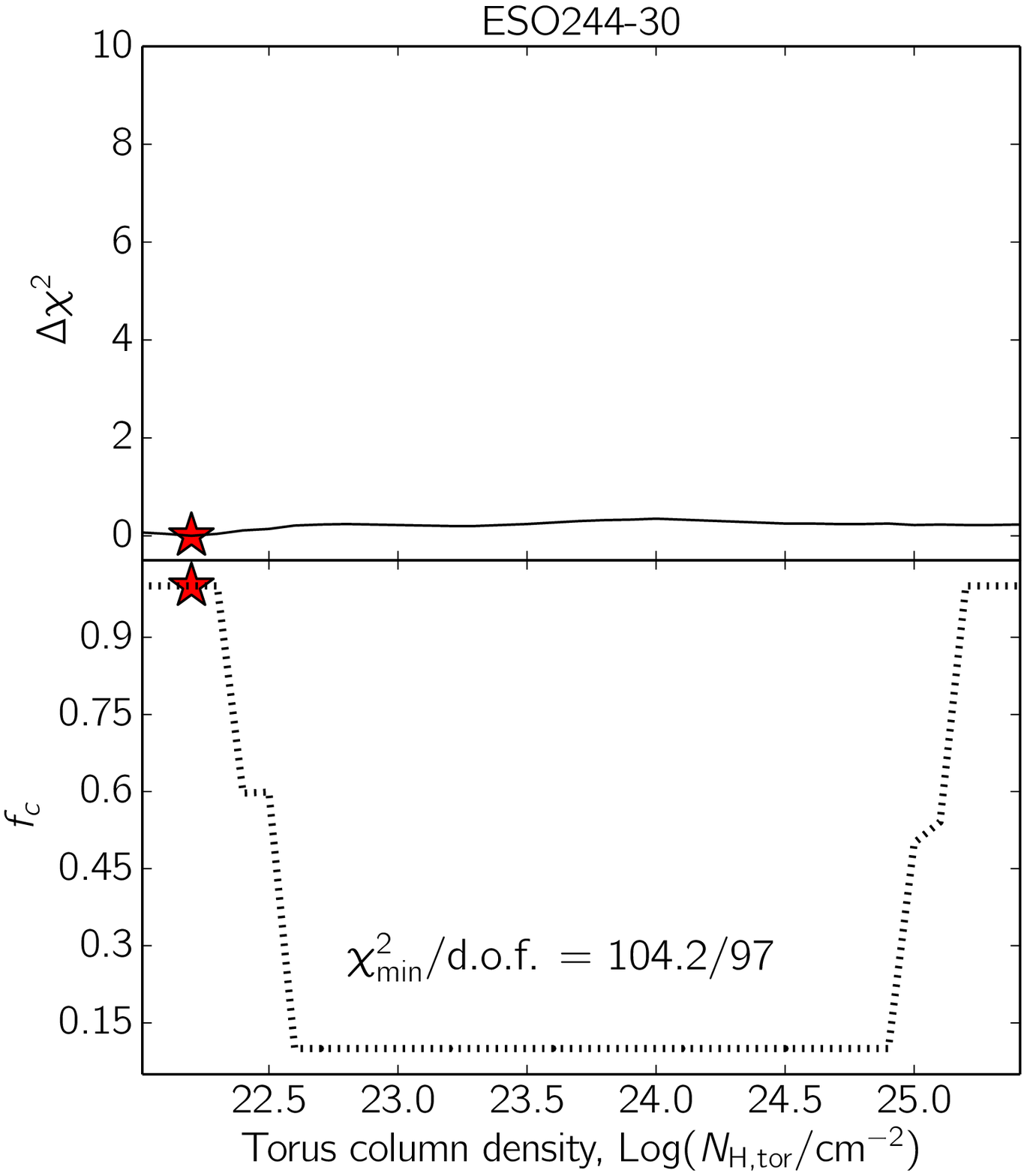}
  \end{minipage}
\begin{minipage}[b]{.33\textwidth}
  \centering
  \includegraphics[width=1.1\textwidth]{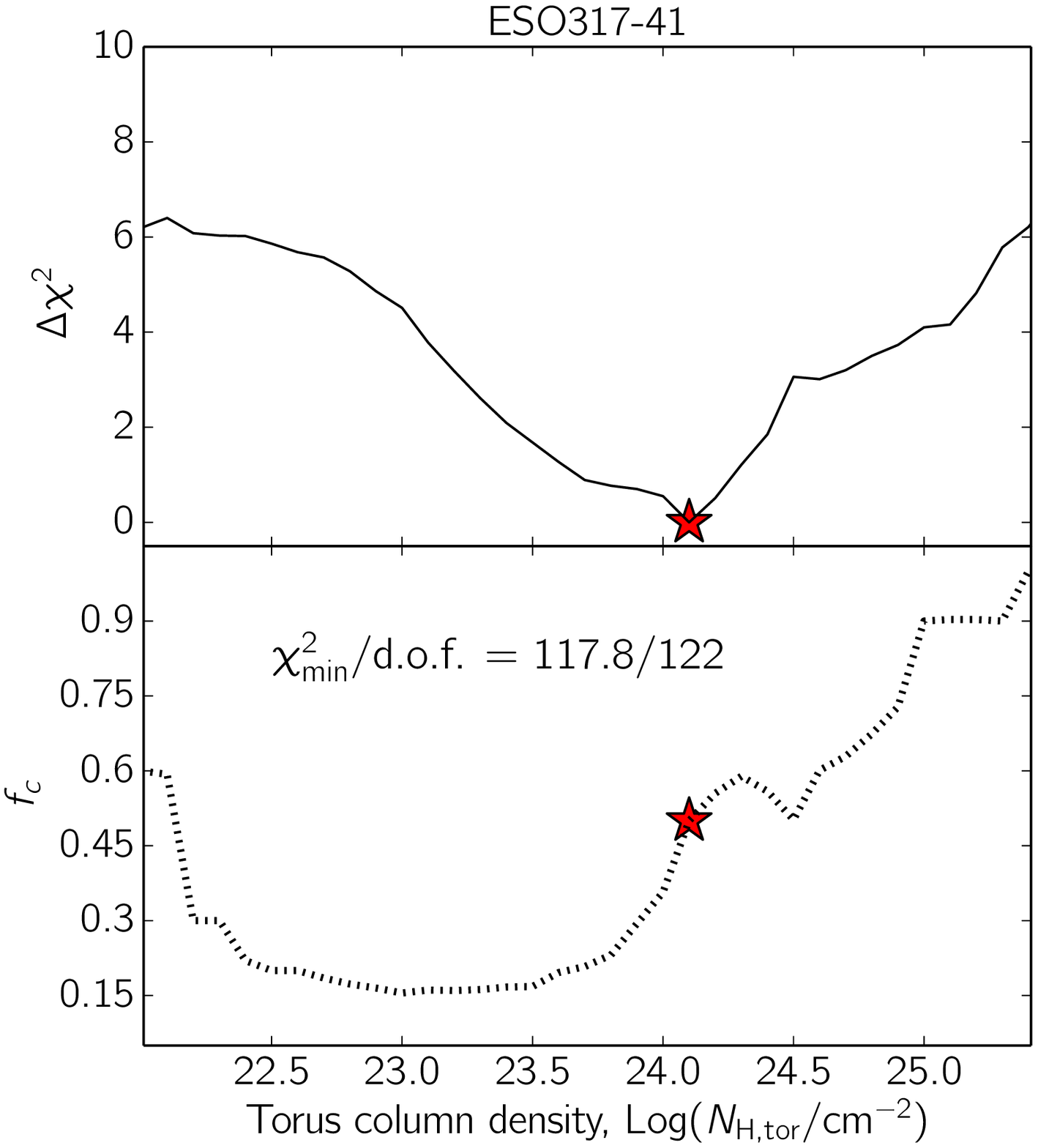}
  \end{minipage}
  \begin{minipage}[b]{.33\textwidth}
  \centering
  \includegraphics[width=1.12\textwidth]{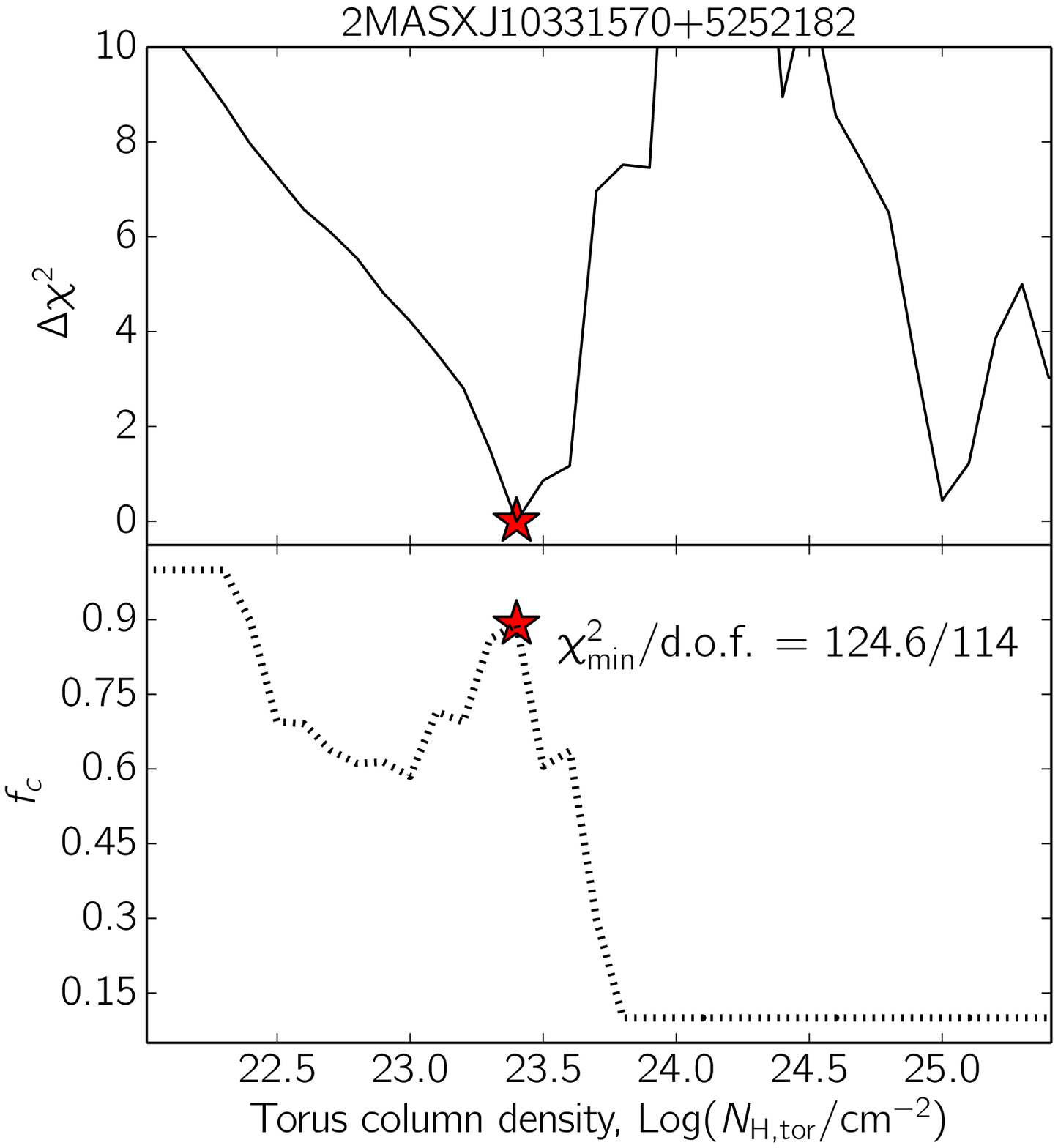}
    \end{minipage}
  \caption{\small Estimate of the torus covering factor in ESO 244-IG 030 (left) and ESO 317-G 041  (center) and  2MASX J10331570+5252182. Top panel: difference between the best-fit, minimum $\chi^2$ ($\chi^2_{\rm min}$) and the $\chi^2$ associated to $\log$($N_{\rm H, tor}$), as a function of the torus average column density. Bottom panel: torus covering factor as a function of the torus average column density. In both panels, we plot as a red star the combination of parameters associated to $\chi^2_{\rm min}$.}\label{fig:logNH_vs_chi_cf}
\end{figure*}

To give a better sense of how well $f_c$ and N$\rm _{H,tor}$ are constrained in each source, in Figure \ref{fig:logNH_vs_chi_cf} we report how both $f_c$ and the best-fit $\chi^2$ vary as a function of the torus average covering factor. As can be seen, in ESO 244-IG 030 (left panel) the fit is insensitive to variations in N$\rm _{H,tor}$ or $f_c$, and the difference in $\chi^2$ between a fit with Log(N$\rm _{H,tor}$)=22.2 and $f_c$=1.0 and one with Log(N$\rm _{H,tor}$)=25.0 and $f_c$=0.1 is $\Delta$$\chi^2$$<$1. In ESO 317-G 041 (central panel) the covering factor is loosely constrained, since for the majority of average column density values (i.e., for Log(N$\rm _{H,tor}$)$<$24.6) we find that the best fit $f_c$ is $f_c$$<$0.6. 
Finally, as already discussed in Section \ref{sec:J1033}, there are two potential solutions for 2MASX J10331570+5252182 (right panel of Figure \ref{fig:logNH_vs_chi_cf}): the one we favor based on the physical reliability of the best fit parameters is the high-$f_c$ one,  ($f_c$=0.90$_{-0.24}^{+0.10}$ for an average torus column density Log(N$\rm _{H,tor}$)=23.4). The low--$f_c$ solution ($f_c$=0.1; Log(N$\rm _{H,tor}$)=25) instead, while statistically equivalent ($\Delta$$\chi^2$=0.3), is ruled out because of the extreme AGN luminosity (Log(L$_{2-10}$)$>$46) implied by the model.
 
\subsection{Conclusions}
We analyzed the joint \nus\ and \xmm\ spectrum of three candidate CT-AGNs selected in the 100-month BAT catalog: these objects were previously classified as CT--AGNs based on low quality \xrt\ and \swi\ data, and had soft ($\Gamma$$>$2.2) photon indices that suggested a potential overestimation of the line of sight column density ($N_{\rm H, l.o.s.}$). In all three sources, we found that both $\Gamma$ and $N_{\rm H, l.o.s.}$ were indeed significantly overestimated: for two objects (ESO 244-IG 030 and ESO 317-G 041) this also implied a reclassification of the source from Compton thick to Compton thin, thus confirming a trend already observed in previous works \citep{marchesi18,marchesi19}, which affects the local Universe intrinsic CT fraction measurement and, consequently, the predictions of the AGN population synthesis models \citep[e.g.,][]{ananna19}.

Thanks to the good \xmm\ spatial resolution and the excellent quality of the \nus\ and \xmm\ spectra, we find that the source 4BPCJ1033.4+5252 \citep[i.e., source SWIFTJ1033.8+5257 in ][]{ricci15} was mistakenly associated to the galaxy SDSS J103315.71+525217.8 at $z$=0.0653: the correct counterpart is instead 2MASX J10331570+5252182. This source redshift ($z$=0.14036) makes it one of the most distant candidate CT-AGN detected by \swi, and possibly even a rare CT quasar.

Finally, the high quality of the \nus\ and \xmm\ data allowed us to constrain several parameters that were previously unconstrained, particularly in the two more obscured objects, i.e., ESO 317-G 041 and 2MASX J10331570+5252182: we measured the Iron K$\alpha$ equivalent width, the torus covering factor and the average torus column density. We found that in ESO 317-G 041 the obscuring material is fairly uniformly distributed, while in 2MASX J10331570+5252182 the l.o.s. column density is significantly higher than the torus average column density, which hints to a possible ``patchy torus'' scenario.

\subsection*{Acknowledgements}
S.M., M.A., and X.Z. acknowledge funding under NASA contract 80NSSC17K0635 and 80NSSC19K0531. A.C. acknowledges support from the ASI/INAF grant I/037/12/0-011/13.

\bibliographystyle{aa}
\bibliography{Marchesi_CTAGN_soft_Gamma}

\appendix
\section{Spectral analysis of 2MASX J10313591--4206093, a source serendipitously detected in the field of view of ESO 317--G041}
In this Appendix we report the joint \nus--\xmm\ spectral analysis of 2MASX J10313591--4206093, which we detected in the field of view of ESO 317--G041 in both the \nus\ and the \xmm\ observation: 2MASX J10313591--4206093 is located at a distance of $\sim$205$^{\prime\prime}$\footnote{The 95\,\% confidence error radius of 4PBC J1031.5--4203, the \swi\ source associated to ESO 317--G041, is $r$=119$^{\prime\prime}$, thus we do not expect any significant contribution of 2MASX J10313591--4206093 to the 15--150\,keV spectrum of 4PBC J1031.5--4203.} from ESO 317--G041, and its redshift is $z$=0.06112. In Table \ref{tab:results_serendip} we report the best-fit parameters obtained in our analysis.

Following the same approach used to analyze the three objects previously discussed in this work, we first fit the \nus--\xmm\ spectrum with \myt\ used in its ``coupled'' configuration. The best fit statistic is good ($\chi^2$/d.o.f.=147.5/144), the source has a relatively hard best-fit photon index ($\Gamma$=1.49$_{-0.09}^{+0.16}$) and is moderately obscured, having Log(N$\rm _{H,l.o.s.}$)=23.54$_{-0.06}^{+0.06}$. 

\begin{figure*}
\begin{minipage}[b]{.5\textwidth}
  \centering
  \includegraphics[width=0.78\textwidth,angle=-90]{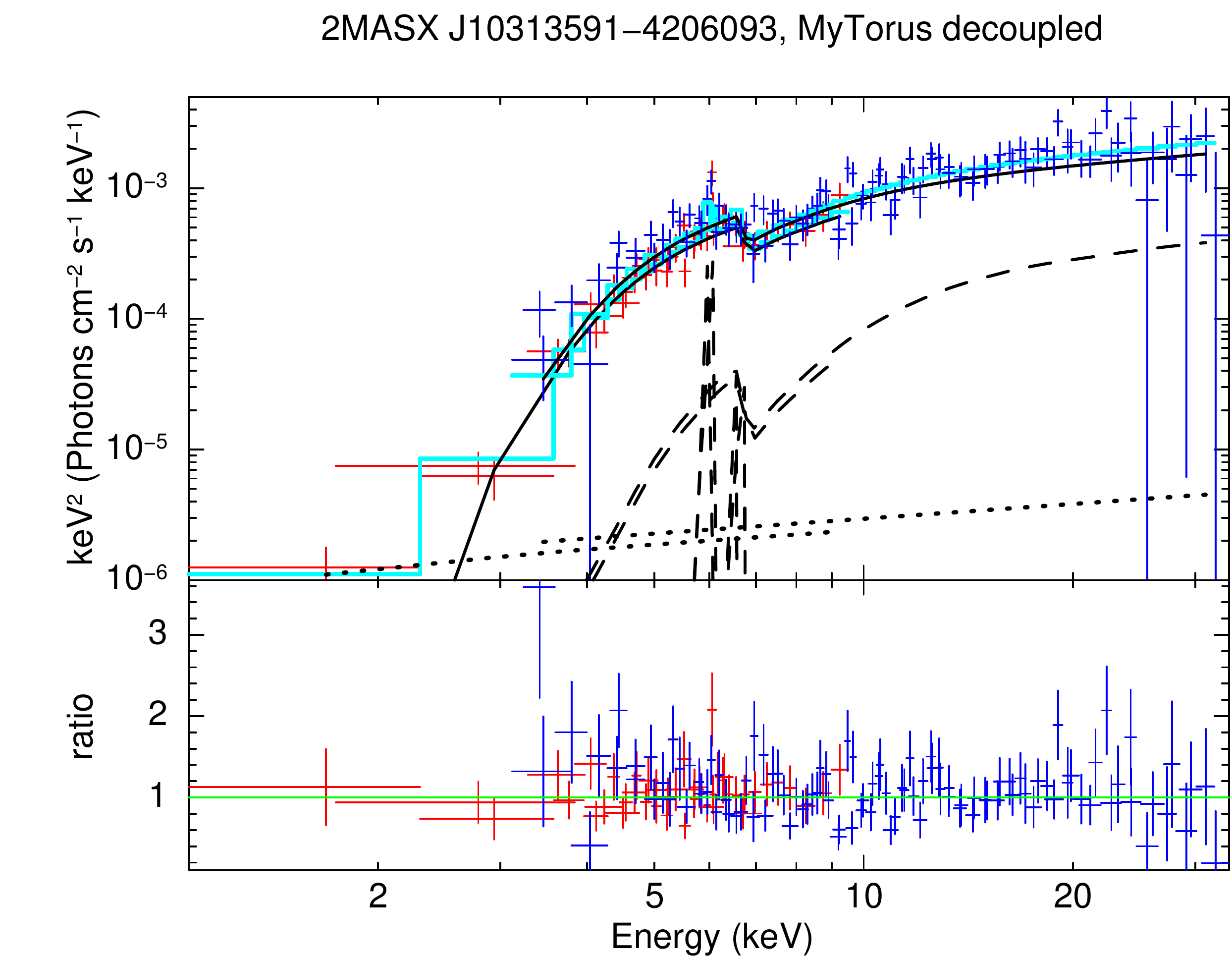}
  \end{minipage}
\begin{minipage}[b]{.5\textwidth}
  \centering
  \includegraphics[width=0.78\textwidth,angle=-90]{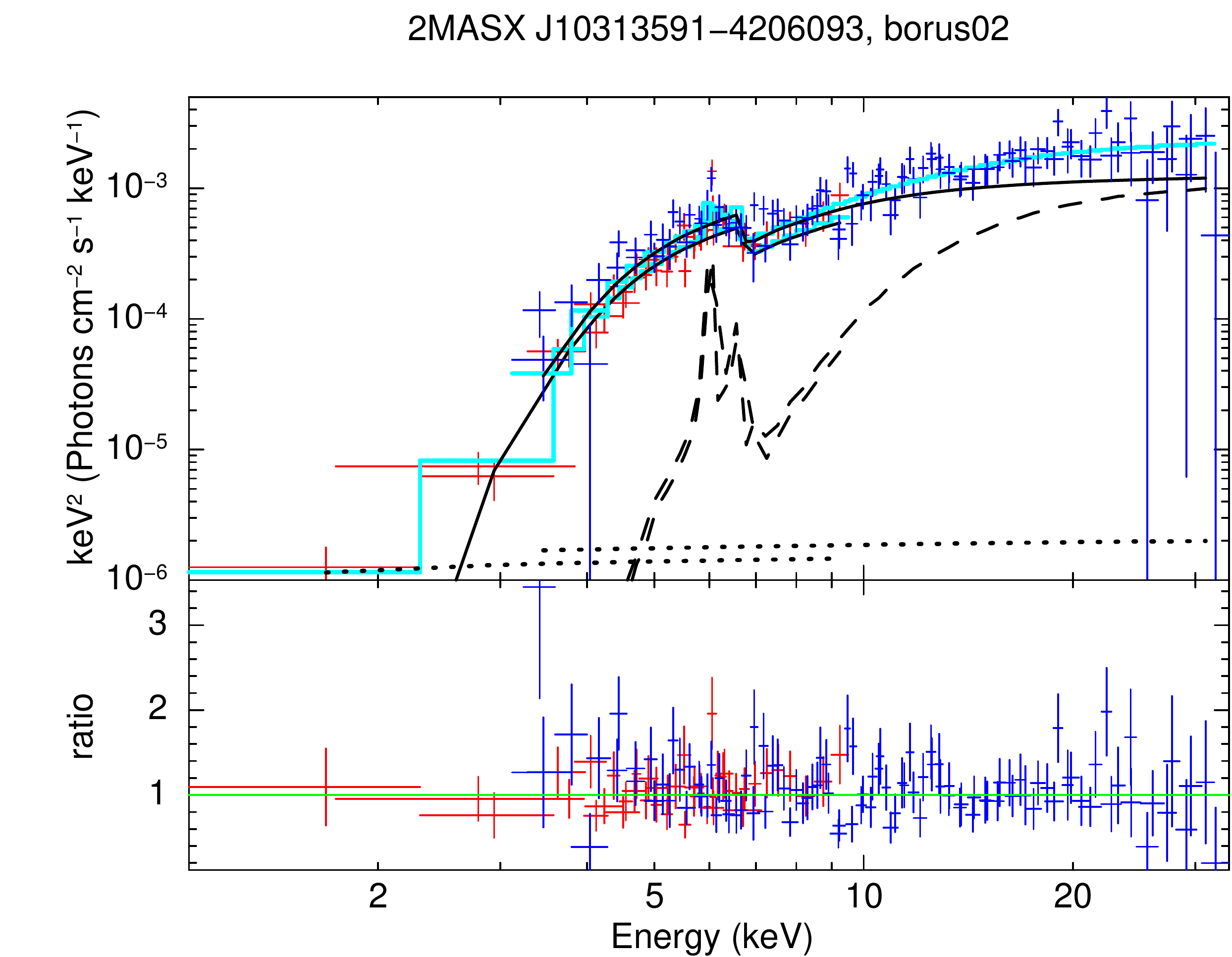}
  \end{minipage}
\caption{\small Background-subtracted spectra (top panel) and data-to-model ratio (bottom) of 2MASX J10313591--4206093. \xmm\ data are plotted in red, \nustar\ data in blue. In the left column we report the best-fit models obtained using \myt\ in ``decoupled'' configuration, while in the right column we show the \borus\ best-fit models. The best-fitting model is plotted as a cyan solid line, the main emission component is plotted as a black solid line, the reprocessed component is plotted as a black dashed line and the main power law component scattered, rather than absorbed, by the torus is plotted as a black dotted line.}\label{fig:spectra_serendip}
\end{figure*}

Using \myt\ in its ``decoupled configuration'' (i.e., allowing the reprocessed component column density, N$\rm _{H,tor}$, to vary independently from N$\rm _{H,l.o.s.}$) instead, leads to a slightly softer solution for the main power law photon index ($\Gamma$=1.63$_{-0.23}^{+0.20}$). The line of sight column density is consistent with the one measured with ``\myt\ coupled'', being Log(N$\rm _{H,l.o.s.}$)=23.55$_{-0.06}^{+0.07}$, while the column density responsible for the reprocessed emission is slightly higher (N$\rm _{H,tor}$=23.97$_{-0.83}^{+0.45}$), although consistent with the l.o.s. one within the 90\,\% confidence uncertainties.  Finally, the fit statistic ($\chi^2$/d.o.f.=147.5/144) is stastically identical to the one obtained using \myt\ in its coupled configuration. We report the \nus\ and \xmm\ spectra and the ``\myt\ decoupled'' best fit model in Figure \ref{fig:spectra_serendip}, left panel.

Finally, the \borus\ best fit model has a better fit statistic than the two \myt\ models ($\chi^2$/d.o.f.=141.8/144; $\Delta$$\chi^2_{\rm MyT-bor}$=5.7): the main difference between the models is in the best fit photon index, which is significantly softer using \borus\ ($\Gamma$=1.90$_{-0.08}^{+0.14}$), while the l.o.s. column density is in full agreement with those measured using \myt, being Log(N$\rm _{H,l.o.s.}$)=23.58$_{-0.04}^{+0.04}$, and the average torus column density corresponding to the minimum $\chi^2$ (Log(N$\rm _{H,tor}$)=24.2) is consistent with the one measured with ``\myt\ coupled''. Based on the \borus\ results, 2MASX J10313591--4206093 has a large covering factor, $f_c$=1.00$_{-0.37}^{+0.00}$: a large covering factor implies a larger contribution of the reprocessed emission to the overall spectrum at energies above 10\,keV (as shown in Figure \ref{fig:spectra_serendip}, right panel), and explains the softer photon index we find using \borus. Given the better $\chi^2$, we assume that the \borus\ best fit offers the most reliable characterization of 2MASX J10313591--4206093.

In summary, the joint \nus--\xmm\ spectral analysis allows us to reliably claim that 2MASX J10313591--4206093 is a fairly luminous (L$_{2-10}$=1.5--2.5 $\times$ 10$^{43}$\,erg s$^{-1}$), heavily obscured, albeit non-CT AGN.

\begingroup
\renewcommand*{\arraystretch}{1.7}
\begin{table*}
\centering
\vspace{.1cm}
  \begin{tabular}{cccc}
       \hline
       \hline       
        & \myt\ coupled & \myt\ decoupled & \borus \\
       \hline
       $\chi^2$/d.o.f.& 147.5/144 & 147.5/144 & 141.8/144 \\
       $C_{Ins}$ & 1.22$_{-0.11}^{+0.12}$ &  1.21$_{-0.11}^{+0.12}$ & 1.26$_{-0.12}^{+0.13}$\\
       $\Gamma$ & 1.49$_{-0.09l}^{+0.16}$ & 1.63$_{-0.23l}^{+0.20}$ & 1.90$_{-0.08}^{+0.14}$ \\
       norm 10$^{-4}$ & 3.77$_{-0.99}^{+2.26}$ & 5.48$_{-2.48}^{+4.15}$ & 9.68$_{-3.01}^{+4.20}$\\
       $\theta\rm _{obs}$ & 90$^f$ & -- & 87$^f$ \\
       Log(N$\rm _{H,l.o.s.}$) & 23.54$_{-0.06}^{+0.06}$ & 23.55$_{-0.06}^{+0.07}$ & 23.58$_{-0.04}^{+0.04}$\\
       Log(N$\rm _{H,tor}$) & =Log(N$\rm _{H,l.o.s.}$)  & 23.97$_{-0.83}^{+0.45}$  &  24.2$^f$\\
       A$_S$ & 1.35$_{-0.81}^{+1.06}$ & 1.00$^f$  & -- \\
       $f_c$ & 0.50$^f$ & -- & 1.00$_{-0.37}^{+0.00u}$\\
       $f_s$ 10$^{-2}$& 0.2$_{-0.1}^{+0.3}$ & 0.2$_{-0.1}^{+0.2}$ & $<$0.2\\
       F$_{2-10}$ & 6.14$_{-1.37}^{+0.21}$ & 6.13$_{-3.51}^{+0.32}$ & 5.88$_{-1.36}^{+0.32}$\\
       F$_{15-55}$ & 46.71$_{-13.75}^{+3.15}$ &  44.56$_{-18.72}^{+3.12}$ & 42.94$_{-15.01}^{+9.79}$\\
       Log(L$_{2-10}$) & 43.25$_{-0.75}^{+0.26}$ & 43.33$_{-0.67}^{+0.25}$ & 43.40$_{-0.64}^{+0.25}$\\
       Log(L$_{15-55}$) & 43.64$_{-0.66}^{+0.24}$ & 43.48$_{-0.79}^{+0.24}$ & 43.24$_{-0.44}^{+0.23}$\\
        \hline
	\hline
	\vspace{0.02cm}
\end{tabular}
\caption{\small \raggedright Best-fits results for the joint \nus--\xmm\ spectral fitting of 2MASX J10313591--4206093. $C_{Ins}$ = $C_{NuS/XMM}$ is the cross calibration between \NuSTAR\ and \XMM, $z$ is the source redshift, $\Gamma$ is the power law photon index, norm is the main power law normalization at 1\,keV, in units of photons s$^{-1}$ cm$^{-2}$ keV$^{-1}$, N$\rm _{H,eq}$ is the equatorial column density, in units of cm$^{-2}$,  $\theta\rm _{obs}$ is the inclination angle between the observer line of sight and the torus axis, A$_S$ is the intensity of the reprocessed component, $f_s$ is the fraction of emission scattered, rather than absorbed by the obscuring torus, $kT$ is the temperature (in keV) of the phenomenological \texttt{mekal} component used to fit the excess at $<$1\,keV. N$\rm _{H,l.o.s.}$ is the line of sight column density, while N$\rm _{H,S}$ and N$\rm _{H,tor}$ are the average torus column densities as measured using \myt\ decoupled and \borus, respectively, and $f_c$ is the torus covering factor, $f_c$ = cos($\theta_{\rm tor}$). Finally, F$_{2-10}$ and F$_{15-55}$ are the observed fluxes (in units of $10^{-13}$ erg cm$^{-2}$ s$^{-1}$) in the 2--10\,keV and 15--55\,keV bands, while L$_{2-10}$ and L$_{15-55}$ are the intrinsic luminosities (in units of erg s$^{-1}$) in the same bands.}
\label{tab:results_serendip}
\end{table*}
\endgroup

\end{document}